\documentclass[aps,prx,floatfix,reprint,superscriptaddress,nobibnotes,showkeys]{revtex4-2}
\usepackage[latin9]{inputenc}
\usepackage{float}
\usepackage{units}
\usepackage{siunitx}
\usepackage{hyperref}
\usepackage[toc,page]{appendix}
\hypersetup{pdftitle={EMH PRL},
	    	pdfauthor={Amirhossein Molavi Tabrizi, Northeastern University},
	    	colorlinks,
	    	pdfcreator={pdflatex},
	    	unicode=false,
	    	pdftoolbar=false,
	    	pdfmenubar=true,
	   		pdffitwindow=true,
	   		pdfnewwindow=true,
	    	linkcolor=red,
	    	citecolor=red,
	    	filecolor=black,
	    	urlcolor=blue,
	    	}
\usepackage{mathrsfs}
\usepackage{amsthm}
\usepackage{bm}
\usepackage{amsmath}
\usepackage{amssymb}
\usepackage{graphicx}
\usepackage{pbox}
\usepackage{esint}
\usepackage{nomencl}
\usepackage{diagbox,ragged2e}
\usepackage{algpseudocode}

\usepackage{tabularx}
\usepackage{booktabs}

 \graphicspath{{../}}

\makeatletter
\usepackage[normalem]{ulem}

\floatstyle{ruled}

\newfloat{algorithm}{tbp}{loa}
\floatname{algorithm}{Algorithm}
\theoremstyle{plain}

\theoremstyle{definition}

\theoremstyle{remark}

\usepackage{amsxtra}
\usepackage{pdfsync}
\usepackage{physics}
\usepackage{xargs}                      
\usepackage[pdftex,dvipsnames]{xcolor}
\usepackage[colorinlistoftodos,prependcaption,textsize=tiny]{todonotes}
\newcommandx{\unsure}[2][1=]{\todo[linecolor=red,backgroundcolor=red!25,bordercolor=red,#1]{#2}}
\newcommandx{\change}[2][1=]{\todo[linecolor=blue,backgroundcolor=blue!25,bordercolor=blue,#1]{#2}}
\newcommandx{\info}[2][1=]{\todo[linecolor=OliveGreen,backgroundcolor=OliveGreen!25,bordercolor=OliveGreen,#1]{#2}}
\newcommandx{\improvement}[2][1=]{\todo[linecolor=Plum,backgroundcolor=Plum!25,bordercolor=Plum,#1]{#2}}
\newcommandx{\thiswillnotshow}[2][1=]{\todo[disable,#1]{#2}}

\newcommand{\ie}{\emph{i.e., }}
\newcommand{\eg}{\emph{e.g., }}

\newcommandx{\AM}[1]{\textcolor{black}{#1}}
\newcommandx{\AMT}[1]{\textcolor{black}{#1}}
\newcommandx{\AK}[1]{\textcolor{black}{#1}}
\newcommandx{\JC}[1]{\textcolor{black}{#1}}
\newcommandx{\Rev}[1]{\textcolor{black}{#1}}
\newcommand\redsout{\bgroup\markoverwith{\textcolor{red}{\rule[0.5ex]{2pt}{1pt}}}\ULon}

\makeatother
  \providecommand{\definitionname}{Definition}
  \providecommand{\remarkname}{Remark}

\begin{document}

\title{Spatiotemporal Organization of Electromechanical Phase Singularities during High-Frequency Cardiac Arrhythmias}

\author{A.~Molavi Tabrizi}
\email{a.molavitabrizi@northeastern.edu}

\author{A.~Mesgarnejad}
\email{a.mesgarnejad@northeastern.edu}

\affiliation{Center for Inter-disciplinary Research on Complex Systems, Department of Physics, Northeastern University, Boston, MA. 
02115, U.S.A.}

\author{M. Bazzi}
\email{bazzim@summahealth.org}
\affiliation{Summa Health System, Akron, OH, USA}

\author{S.~Luther}
\email{stefan.luther@ds.mpg.de}
\affiliation{Max Planck Institute for Dynamics and Self-Organization, G\"{o}ttingen, Germany}
\affiliation{German Center for Cardiovascular Research (DZHK), partner-site Goettingen, Germany}

\author{J.~Christoph}
\email{jan.christoph@ucsf.edu}
\affiliation{Cardiovascular Research Institute, University of California, San Francisco, CA, USA}

\author{A.~Karma}
\email{corresponding author: a.karma@northeastern.edu}

\affiliation{Center for Inter-disciplinary Research on Complex Systems, Department of Physics, Northeastern University, Boston, MA. 
02115, U.S.A.}

\date{\today}

\begin{abstract}

Ventricular fibrillation (VF) is a life-threatening electromechanical dysfunction of the heart associated with complex spatiotemporal dynamics of electrical excitation and mechanical contraction of the heart muscle. 
It has been hypothesized that VF is driven by three-dimensional rotating electrical scroll waves, which can be characterized by filament-like electrical phase singularities or vortex filaments, but visualizing their dynamics has been a long-standing challenge. 
Recently, it was shown that rotating excitation waves during VF are associated with rotating waves of mechanical deformation.
Three-dimensional mechanical scroll waves and mechanical filaments describing their rotational core regions were observed in the ventricles by using high-resolution ultrasound. The findings suggest that the spatiotemporal organization of cardiac fibrillation may be assessed from waves of mechanical deformation. 
However, the complex relationship between excitation and mechanical waves during VF is currently not understood.
Here, we study the fundamental nature of mechanical phase singularities, their spatiotemporal organization and relation with electrical phase singularities. 
We demonstrate the existence of two fundamental types of mechanical phase singularities: ``paired singularities'', which are co-localized with electrical phase singularities, and ``unpaired singularities'', which can form independently.
We show that the unpaired singularities emerge due to the anisotropy of the active force field, generated by fiber anisotropy in cardiac tissue, and the non-locality of elastic interactions, which jointly induce strong spatiotemporal inhomogeneities in the strain fields. 
The inhomogeneities lead to the breakup of deformation waves and create mechanical phase singularities, even in the absence of electrical singularities, which are typically associated with excitation wave break. 
We exploit these insights to develop an approach to discriminate paired and unpaired mechanical phase singularities, which could potentially be used to locate electrical rotor cores from a mechanical measurement. Our findings provide a fundamental understanding of the complex spatiotemporal organization of electromechanical waves in the heart, and a theoretical basis for the analysis of high-resolution ultrasound data for the three-dimensional mapping of heart rhythm disorders.

\end{abstract}

\maketitle

\section{Introduction}

The beating of the heart is initiated by nonlinear waves of electrical excitation, which propagate through the cardiac muscle and trigger a release of intracellular calcium, which in turn triggers contractions of cardiac muscle cells.
During severe heart rhythm disorders, such as atrial or ventricular fibrillation, the electrical excitation degenerates into multiple disorganized, asynchronous electrical waves leading to irregular cardiac muscle contractions \cite{weiss2005dynamics,Karma:2013aa,qu2014nonlinear,Alonso:2016aa}. 
Both spiral-shaped re-entrant waves and focal waves are thought to underlie those phenomena during high-frequency cardiac arrhythmias. 
To date, however, the three-dimensional (3D) electrical wave phenomena that evolve rapidly within the thickness of the heart muscle have never been imaged in full. 
While fluorescence imaging provides high-resolution measurements of vortex-like rotating spiral waves on the heart surface, the underlying 3D dynamics remain elusive, and, based on numerical simulations, are conjectured to take on the shapes of scroll vortex waves or 3D spiral waves \cite{Davidenko1992, Pertsov1993, winfree:1994aa, Jalife1996, Gray1998, Witkowski1998, Qu2007}, which are generic self-organized nonlinear wave structures of excitable media \cite{Welsh1983, Rotermund1990,weiss2005dynamics,Karma:2013aa,qu2014nonlinear,Alonso:2016aa}. 

In a recent study, it was shown that the dynamical processes underlying ventricular fibrillation can be characterized by coupled electrical and mechanical phase singularity dynamics~\cite{Christoph2018}. 
By using tri-modal voltage- and calcium-sensitive fluorescence imaging, high-speed 4D ultrasound, and numerical motion analysis, it was shown that the rapidly contracting, fibrillating heart muscle exhibits vortex-like rotating strain fields, which are produced by electrical action potential and calcium spiral vortex waves.
Consequently, it was demonstrated on the heart surface that electrical phase singularities, which describe the cores of rotating action potential and calcium waves, appear in the vicinity of the cores of the rotating deformation patterns, which can equivalently be characterized by mechanical phase singularities.
It was furthermore shown that electrical and mechanical phase singularities co-localize and exhibit similar dynamics in terms of numbers, trajectories, and life-times, providing for the first time evidence for the existence of electromechanical rotors.
The data suggests that electrical and mechanical phase singularities are both produced during ventricular fibrillation by electromechanical scroll wave chaos, and that a better understanding of the nature of these electromechanical phase singularities could provide important new insights into the 3D spatiotemporal organization of cardiac fibrillation since electrical and mechanical phase singularities are closely related to each other.
Lastly, it was shown using high-resolution 4D ultrasound, that a 3D filament-like structure of mechanical phase singularities evolves throughout the heart wall during fibrillation, suggesting that the spatiotemporal organization of electrical scroll wave vortex filaments may be inferred by the dynamics of mechanical vortex filaments inside the heart muscle.
The findings open the path to use mechanical phase singularities to enhance the understanding of the mechanisms underlying cardiac fibrillation.
However, despite the experimental advances, a fundamental understanding of their relationship remains largely lacking. 

Naively, one would expect mechanical waves to be slaved to electrical excitation waves that cause contraction via the well-established excitation-contraction coupling mechanism, whereby calcium entry into the cell through L-type calcium channels following electrical excitation triggers calcium release from intracellular stores, which in turn activates the contractile machinery of the cell thereby generating an active force along the long axis of cardiomyocytes \cite{Bers2002}. 
Consequently, the spatiotemporal wave pattern of this active force is expected to generally follow closely the wave pattern of electrical excitation, except under extreme conditions where excitation-contraction coupling fails (\eg, when calcium-induced-calcium-release fails to keep up with electrical excitation at very high frequencies). 
\Rev{When normal excitation-contraction coupling remains operative, excitation wave propagation can itself be influenced by mechanical contraction due to mechanoelectrical feedback \cite{kohl2003cardiac,Nash:2004aa,Weise:2013aa}. However, this feedback, which has been neglected in most modeling studies of cardiac arrhythmia mechanisms \cite{weiss2005dynamics,Karma:2013aa,qu2014nonlinear}, does not cause the wave pattern of active force to deviate from the wave pattern of electrical excitation.}
In contrast, the resulting tissue deformation pattern, which is imaged by high-resolution ultrasound, may differ markedly from the active force pattern due to the non-locality of long-range elastic interactions, which can induce strain in regions of the heart muscle that are not actively contracting or, at the opposite, cause regions under large active force to be under small strain. 
This raises the basic question of whether mechanical phase singularities, \ie spatiotemporal phase singularities of non-local strain fields, track electrical wave singularities, or exhibit a more complex spatiotemporal organization. 

In this article, we address this question by investigating the strain field patterns generated by re-entrant and focal excitation-contraction waves in 2D and 3D cardiac tissue in computer simulations.
Our main finding is that mechanical phase singularities, which we extract from phase maps obtained of strain fields via the Hilbert transform, can both co-localize with, or form away from, phase singularities of excitation waves. 
The former ``paired'' mechanical singularities only exist in the presence of re-entrant waves while, remarkably, focal sources of waves or target patterns suffice to create the latter ``unpaired'' mechanical singularities, which can, therefore, exist both in the presence and absence of re-entrant waves. 
Further, we find that paired and unpaired singularities form by different mechanisms. Paired ones originate from phase singularities associated with the rotation of strain fields that remain essentially slaved to excitation waves near the core of electrical vortices. In contrast, unpaired ones originate from the breakup of deformation waves caused by the combination of long-range elastic interactions and the anisotropy of the active force field.  

We exploit these insights to develop an approach to discriminate paired and unpaired mechanical phase singularities based on observations of their surrounding strain field. Our findings provide a fundamental understanding of the complex spatiotemporal organization of electromechanical wave activity during cardiac arrhythmias, and a theoretical basis for the interpretation of three-dimensional imaging data of tachyarrhythmias obtained with high-resolution ultrasound for diagnostic and therapeutic applications.

\section{Methods}\label{sec:methods}

To explore the relationship between electrical and mechanical singularities, we use two different approaches. First, in section~\ref{sec:2d-qstatic}, we analytically construct simple 2D excitation wave patterns and compute the resulting strain fields (deformation waves) by numerically solving the equations of linear elasticity using the finite element method (FEM).
\AM{We should note that there is a large body of literature dedicated to creating accurate representation of the coupled electromechanical response of the heart \cite{Holzapfel:2009aa,Costabal:2017aa,Costabal:2017tp,Sahli-Costabal:2017aa,Propp:2020tk,Nitti:2021th}.
However, since our goal is to elucidate the mechanism of mechanical singularity formation we reduce the problem to its most essential aspects. For the main presentation in this article, we use quasi-static formulation with homogeneous isotropic infinitesimal elasticity, with anisotropic active strains and no electromechanical feedback.
We also limit our presentation in the main text to Poisson's ratio of $\nu=0.4$ which produces $\sim10\%$ volumetric change (see Fig.~\ref{fig:2disopace}).
However, to show the robustness of our findings, we provide an extended set of simulations using $\nu=0.49$, neo-hookean material, and transversely isotropic elasticity in the Supplementary Videos.}
Under the assumption that the local active force tracks the local electrical excitation, which is generally satisfied except when calcium release cannot keep up with very high-frequency electrical waves, we do not need to specify separately the electrical and mechanical signals. 
We prescribe directly the spatiotemporal active \AM{strain} fields corresponding to standard excitation 2D wave patterns. 
As generic examples of reentrant and focal excitations, we consider both a rigidly rotating (\ie non-meandering) Archimedean spiral wave and a target pattern generated by a localized time-periodic focal source of waves, respectively. This approach has the advantage that it provides the simplest possible setting to investigate the relationship of electrical and mechanical singularities in a perfectly elastic, nearly incompressible, material where the strain field is assumed to relax instantaneously to the active force field. In this approach, ``electrical singularities'' are simply the singularities of the prescribed active force fields, which are present and absent for spiral and target waves, respectively, and ``mechanical singularities'' are the singularities of the numerically obtained linear elastic strain fields generated by the active force fields. 

Second, in section~\ref{sec:3dmethods}, we simulate electromechanical wave activity using a simplified two-variable ionic model of electrical excitation \cite{Karma:1993aa,Karma:1994aa} and a simple phenomenological relaxational kinetic equation to relate the active force to electrical excitation \cite{Nash:2004aa}. In addition, we compute the resulting strain fields by extending a mass-spring model  (MSM) of standard elasticity \cite{Kot:2015aa} to incorporate the active force and describe a nearly incompressible viscoelastic material with inertia. This approach has the advantage of allowing us to efficiently explore the relationship of excitation-contraction and deformation waves in 3D anisotropic tissue in a parameter limit of the MSM where viscous and inertial effects are small and continuum elastic properties are similar to those obtained by quasistatic FEM solutions, with known continuum elastic properties that can be derived analytically from the MSM.  

\subsection{Two-dimensional FEM computations with analytically prescribed spiral and target wave active force fields}\label{sec:2d-qstatic}

We perform the computations using quasistatic linear elasticity, where we implemented active contraction by an analogy to thermal stress with the incorporation of (stress-free) eigen strains.
\AM{Our use of active strains ensures the ellipticity of the elastic energy~\cite{Ambrosi:2012tb}}.
For domain $\Omega\subset\mathbb{R}^2$ we write the elastic energy as:
\begin{equation}\label{eq:elastic-energy}
	E_{el}(u)=\int_{\Omega} C_{ijkl}(\epsilon_{kl}+T_a\beta_{kl})(\epsilon_{ij}+T_a\beta_{ij})\,dV
\end{equation}
where $C_{ijkl}=\lambda\delta_{ij}\delta_{kl}+\mu(\delta_{ik}\delta_{jl}+\delta_{il}\delta_{jk})$ is the fourth order isotropic elastic constitutive tensor where $\delta_{ij}$ is the Kronecker delta and for plane-stress $\lambda=E\nu/(1-\nu^2)$ and $\mu=E/(2(1+\nu))$ are Lam\'e parameters, $\epsilon_{ij}=(u_{i,j}+u_{j,i})/2$ is the linearized strain tensor (\ie symmetric displacement gradient), $0\leq T_a \leq 0.15$ is the imposed spatiotemporally varying active contraction field normalized by the elastic modulus $E$, and ${\beta}$ is a tensor whose form given below depends on whether contraction is isotropic, as in the case of randomly oriented cardiomyocytes in a tissue culture, or aniostropic as in the case of heart tissue with aligned fibers.
To obtain the quasistatic response, we find the displacement field $u^*$ that minimizes the elastic energy, Eq.~\eqref{eq:elastic-energy}. 

We assume cardiac tissue to be elastically isotropic in all cases studied here with \ie Young's modulus $E$ and Poisson's ratio $\nu=0.4$ to account for near-incompressibility of the tissue,
\AM{which results in $\sim10\%$ maximum volume change in our FE simulations.
This value of the Poisson ratio was selected to correspond to the value achieved by the MSM model given our choice of parameters (see Appendix~\ref{app:msm_properties}). 
To assess the effect of volume change on our findings, we duplicated our simulations for higher Poisson's ratio $\nu=0.49$ (See Supplementay Movies).}
We investigate two different 2D cases:

\noindent \textbf{Isotropic case}. 
 
The ``isotropic'' case mimics a thin quasi-2D tissue culture of randomly oriented cardiomyocytes with no preferred direction of conduction or active contraction.
In this case, both the conduction and active strains are assumed to be isotropic.
Furthermore, we set
\begin{equation}\label{eq:isotropic-vegard}
 	\beta_{ij}=\delta_{ij}
 \end{equation} 
to produce an isotropic contraction of unit magnitude.

\noindent \textbf{Anisotropic case}.
 
The ``anisotropic'' case represents normal 2D cardiac tissue with cardiomyocytes aligned along a common fiber axis where we assume the conduction along the fibers to be 5 times faster than conduction perpendicular to them.
We further assume that the active strain only acts along the fiber.
In this case, we choose the expansion tensor such that for a 1D fiber, the strains along the fiber converge to unit magnitude without any change of fiber cross-section.
For a domain with fibers along the first coordinate axis, and assuming isotropic elasticity, this is achieved by choosing
\begin{equation}\label{eq:anisotropic-vegard}
	\beta=\begin{bmatrix}
	 -1 &  0 \\
	 0 & \nu  \\
	 \end{bmatrix}
\end{equation}
We note that the above equations \eqref{eq:isotropic-vegard} and \eqref{eq:anisotropic-vegard} are special cases of a domain with arbitrary fiber orientation defined by the unit vector $f$ with components $(f_1,f_2)$ with $\beta$ defined by
\begin{equation}\label{eq:eigen-strains}
	\frac{C_{ijkl}}{\lambda+2\mu}\,\beta_{kl}=-f_{i}f_{j}
\end{equation}

To perform the simulation, we explicitly impose the active contraction field $T_a$ using closed-form expressions in the form of spatially-diffuse Archimedean spirals and expanding ellipsoidal pulses (see Appendix~\ref{app:spiral}) in an $L\times L$ square domain discretized using $320\times320$ Q1 elements.
We impose traction-free boundary conditions on the edges of the domain ($x=\pm L$ and $y=\pm L$) and remove the rigid body modes (null-space of elasticity) from the obtained discrete set of equations. 
For our FEM numerical implementation, we use libMesh~\cite{Kirk:2006a} for finite element book-keeping and PETSc~\cite{petsc-user-ref,petsc-efficient} for linear algebra. 

\subsection{Three-dimensional simulations using a two-variable ionic model with active force generation coupled to a mass-spring model}\label{sec:3dmethods}
\subsubsection{Reaction diffusion model for cardiac excitation}\label{sec:karma94}

In 3D simulations, we assume a slab of ventricular muscle with a defined fiber architecture.
A two-variable reaction-diffusion model presented in \cite{Karma:1993aa,Karma:1994aa} has been utilized to simulate the cardiac excitation. The magnitude of the active force is coupled phenomenologically to the voltage field to mimic the typical rise and fall of this force during an action potential \cite{Nash:2004aa}. 
The equations that describe this model are:
\begin{align}
	&\partial_{t}U = \nabla\cdot(\mathbf{D}\nabla U)+\tau^{-1}_{U}f(U,v)\label{eq:uequation}\\
	&\partial_{t}v = \tau^{-1}_{v}g(U,v)\label{eq:vequation}\\
	&\partial_{t}T_a = \chi(U)(K_tU-T_a)\label{eq:taequation}
\end{align}
subjected to the Neumann boundary condition:
\begin{align}
	\hat{n}\cdot(\mathbf{D}\nabla U)=0.\label{eq:bcequation}
\end{align}
In these equations, $U$ and $v$ are the dimensionless representation of the transmembrane voltage and slow current gate variables, respectively, $\tau_U/\tau_v$ controls the relative abruptness of the excitation, $\mathbf{D}$ is the diffusivity tensor, and $\hat{n}$ is the unit normal vector to the boundary. 
In Eq.~\eqref{eq:taequation} (similar to Eq.~\eqref{eq:elastic-energy}), $T_a$ represents the ratio of the active contraction field and elastic modulus.
$K_t$ controls the maximum amplitude of the active contraction by coupling it to the trans-membrane voltage $U$, and $\chi(U)$ is a step function that sets the time scale of the contraction period with respect to $U$. 
Details about functions $f$, $g$, and the relation between diffusivity tensor and the fiber architecture can be found in \cite{Karma:1994aa,Fenton:1998aa}. 
For convenience, these relations and more details about Eq.~\ref{eq:taequation} are briefly presented in Appendix~\ref{app:RDmodel}.
Finally, with no loss of generality, we assume that the conduction is {transversely isotropic}, \ie it is isotropic in the plane perpendicular to fibers and is faster along the fiber direction.
We can therefore write $D_{\perp1}=D_{\perp2}=D_{\perp}$ where $D_{\perp1}$ and $D_{\perp2}$ are the diffusivity perpendicular to the fiber axis in each plane and $D_\parallel$ is the diffusivity along the fiber axis. 
This will simplify the local electric current to
\begin{align}
\mathbf{D}\nabla U=D_{\perp}\nabla U+(D_{\parallel}-D_{\perp})(\hat{f}\cdot\nabla U)
\label{eq:localelectriccurrent}
\end{align}
where $\hat{f}$ is the unit vector that is locally parallel to the fiber. Here we consider the anisotropic cases without fiber rotation where $\hat{f}$ is spatially uniform and with fiber rotation where $\hat{f}$ rotates along one axis as in \cite{Karma:1994aa,Fenton:1998aa}.
The set of parameters that are used in this article is listed in Table~\ref{Ta:electroparameters}. 

\subsubsection{Three-dimensional lattice mass-spring model}\label{sec:3dlatticemassspringmodel}

\begin{figure}[]
\centering
\includegraphics[width=0.8\columnwidth,angle=0,trim={0.in 0.in 0.in 0.in}]{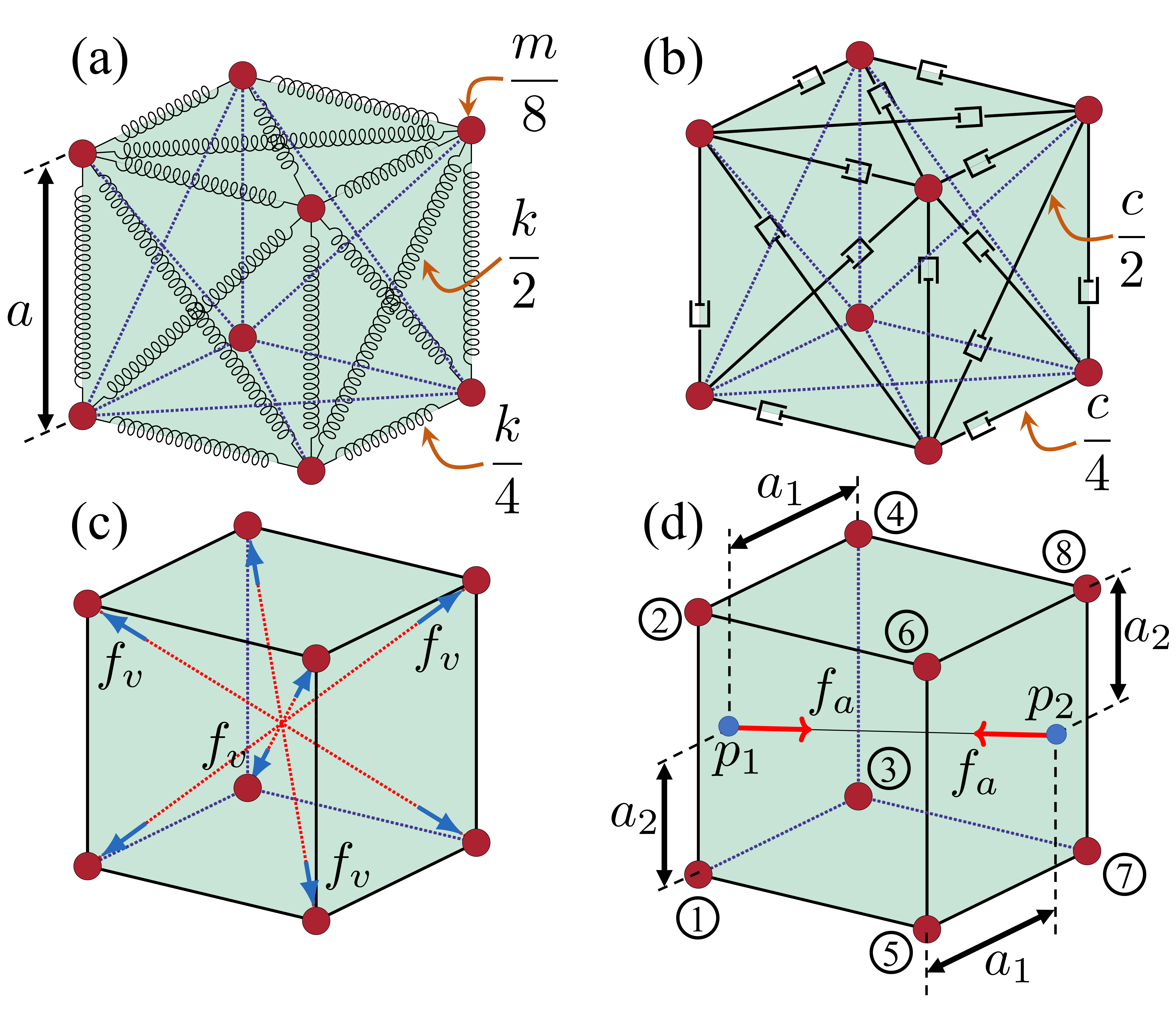}{}
\caption{A cubical element of the mass-spring model (MSM). (a) The mass of the cube $m$ is distributed equally among the 8 cubes that meet at one corner. The edge and diagonal springs have a stiffness of $k/4$ and $k/2$, respectively. (b) A Kelvin-Voigt type damping system is used where the edge and diagonal dampers have damping constant of $c/4$ and $c/2$, respectively. (c) To add volume conservation to the model, a penalty pressure is applied to the faces of the cube. This pressure results in forces at each corner. (d) The compressive active contraction is applied along the fiber direction in each cube and based on the distances $a_1$ and $a_2$ with bilinear interpolation, it has been redistributed on the masses.}
\label{fig:cube}
\end{figure}

In the 3D simulations, the elasticity of the tissue was modeled by extending the mass-spring model (MSM) of Refs.~\cite{Kot:2015aa} to cardiac tissue. 
\AM{Our use of the discrete MSM model (as opposed to discretizing a set of continuum-level equations), allows us to easily integrate the equations of motion on massively parallel GPUs.}
The schematic of a single unit cell of the cubic lattice with edge length $a$ is shown in Fig.~\ref{fig:cube} 
and the tissue is constructed by stacking several unit cells. The extension to cardiac tissue was achieved by 1)~adding Kelvin-type dampers (Fig.~\ref{fig:cube}(b)) to the MSM to account for the viscoelastic behavior of the tissue, which produces damping forces proportional to the relative velocity between two masses, 2)~a volumetric penalty force to shift the elastic properties towards the incompressible limit (Fig.~\ref{fig:cube}(c)), and 3)~by adding the active contractile force (Fig.~\ref{fig:cube}(d)).
For small and slowly spatially varying deformations (\ie varying on a scale much larger than $a$) and in the absence of the volumetric penalty forces, the MSM reduces in the continuum limit to standard isotropic linear elasticity with Lam\`e constants $\lambda=\mu=k/a$ characteristic of a compressible material \cite{Kot:2015aa}.

However, similar to other biological tissues, the myocardium contains mostly water and can be considered as a {\it nearly}  incompressible material \cite{Holzapfel:2009aa}. 
To account for this near-incompressibility in our model, we introduce a uniform penalty pressure $p$ to keep the volume constant.
This pressure is applied to all faces of a unit cell and transmits a force on each mass at the corners of the cell. 
A schematic of the volumetric forces for the case that the volume decreases is shown in Fig.~\ref{fig:cube}(c). 
The magnitude of the volumetric penalty force is chosen as 
\begin{align}
\label{eq:volumetric_magnitude}
f_v= \frac{\sqrt{3}~p a^2}{4}\left(1-\frac{V}{V_0}\right)
\end{align}
where $V_0=a^3$ and $V$ are the initial and current volume of each cubical element, respectively. 
To ensure the balance of the internal forces, we apply the volumetric forces on each pair of diagonally opposite nodes along the line connecting them.

This volume preservation method can be shown to modify the bulk stiffness of the MSM. 
In the presence of the volumetric pressure, $p$, the first Lam\'e constant is increased to $\lambda=k/a+p$ while the shear modulus remains unchanged $\mu=k/a$.
Therefore we can write Young's modulus and Poisson's ratio of the system as:
\begin{align}
	E=\frac{k(3ap +5k)}{a(ap +2k)}\label{eq:youngvol}\\
	\nu=\frac{ap +k}{2 a p +4k}\label{eq:nuvol}
\end{align}
In the simulations, we chose $p=15k/(2a)$ corresponding to a Poisson's ratio $\nu=17/38\approx 0.45$ close to the incompressible limit.
\AM{We should note that while the value of penalty forces~\eqref{eq:volumetric_magnitude} vary spatially, the underlying model remains isotropic and homogeneous.
}

Next, \AM{we incorporate the active contraction in our model. The active tension can be incorporated in two ways: (i) using active strains by dynamically setting the rest length of passive strains or (ii) using active stresses.
Due to its ease of use in the MSM model, we opt for the latter option. We should note that these two options are equivalent only at small strains and can produce quantitatively different results at finite strains \cite{Ambrosi:2012tb}.}
We assume that contraction only occurs along the local fiber axis. 
In each unit cell, we assume that there is a pair of internal active forces oriented in the direction of the fiber and choose the magnitude of these forces as $|f_a|=\bar{T}_a E a^2$ where $\bar{T}_a=\sum_{i=1}^{8} T_a^{i^{th} cell}/8$ represents the average contractile field in the cell.
The active force contribution from each cubical cell $T_a^{i^{th} cell}$ at a corner is calculated using bilinear interpolation based on the position where the fiber axis intersects the plane containing the mass, as described previously \cite{Bourguignon2000, Lebert2019}, and shown in Fig.~\ref{fig:cube}(d). 
Finally, the equation of motion for the $i_{th}$ mass is given by 
\begin{align}
\label{eq:motion}
m\partial^{2}_{t}u_i=\sum_{j=1}^{18}f_s+\sum_{j=1}^{18}f_d+\sum_{j=1}^{8}f_a+\sum_{j=1}^{8}f_v
\end{align}
where the various forces shown are in Fig.~\ref{fig:cube}. They include the static spring forces $f_s=k\times$(relative displacements of masses linked by springs) and associated damping forces $f_d=c \times$(relative velocities) for the 18 dampers connected to each mass, and the active forces $f_a$ and volumetric forces $f_v$ defined above for the 8 cubical elements connected to each mass. 
\AMT{A detailed study of the mechanical properties of the MSM has been conducted and presented in Appendix~\ref{app:msm_properties}}.

To study the electrical and mechanical singularities in heart tissue, we perform the simulations on a cubic lattice with $150\times150\times40$ grid points. 
Adjusting for the material properties, the lattice corresponds to a $4.5\times4.5\times1.2\,\mathrm{cm}^3$ slab with a unit cell spacing $\Delta x=\Delta y=\Delta z=a=0.3\,\mathrm{mm}$.
To integrate the electrophysiology equations (Eqs.~\eqref{eq:uequation}--\eqref{eq:taequation}) we use an explicit Euler method.
To create a spiral wave initial condition, we use the standard two-stimulus protocol that consists of first creating a traveling plane wave along the $y$ direction by exciting the tissue uniformly on the $y=0$ plane, and then depolarizing half of the tissue ($x<2.25\,\mathrm{cm}$) at $t=0.35\,\mathrm{s}$. 
In addition, we integrate the equations of motion of the MSM using a standard velocity Verlet algorithm.

In a human heart with a density close to water $\rho\approx 1000\,\mathrm{kgm}^{-3}$ and Young's modulus $\approx 125\,\mathrm{MPa}$~\cite{Weise:2013aa}, the elastic wave speed can be crudely estimated to be $C\approx 353\,\mathrm{ms^{-1}}$. 
Furthermore, the period of excitation $T$ spans the range $0.1$ to $1.0\mathrm{s}$ with the lower and upper bounds corresponding to an electrical rotor and a normal heartbeat, respectively. 
It follows that the distance $CT$ traveled by elastic waves during a contraction period is much larger than the characteristic size $L$ of the heart of a few centimeters. 
This scale separation ($L\ll CT$) makes it possible to increase the computational efficiency of the time integration algorithm by reducing the elastic wave speed.
Therefore, instead of setting Young's modulus to its physical value, we set the characteristic elastic wave speed $C=\sqrt{E/\rho}$ equal to $5\,\mathrm{ms}^{-1}$ such that the inequality $L \ll CT$ still holds, where $\rho=m/a^3$ is the tissue density.

\AM{The passive response of cardiac myocyte exhibits hysteresis cycles indicative of a viscoelastic behavior \cite{Holzapfel:2009}.
While these effects can create small quantitative differences compared to the non-viscoelastic models there is no evidence that they can significantly change the response of the tissue which is primarily driven by the active stresses~\cite{Propp:2020tk}. As such in our implementation we neglect such effects and} we choose the damping constant such that the length of the system is much smaller than the spatial decay length of linear elastic waves, or $L\ll (CT)/(\pi \delta)$, where 
$\tan(\delta)=E_l/E_s$ with $E_l$ and $E_s$ are the loss and storage moduli, respectively, of our viscoelastic medium.
In addition, we choose the elastic diffusion constant $D_e = \eta/\rho=11~ \mathrm{cm}^{2}s^{-1}$  where $\eta=5c/(2a)$ is the Kelvin-Voigt damping constant.
We present a more detailed justification for these choices of parameters in Appendix~\ref{app:1dcable}.
The parameters for the MSM model are summarized in Table~\ref{Ta:electroparameters}.
Fig.~\ref{fig:cube} implies that the sum of moments and forces in each cube and consequently in the full system is always zero.
Therefore, all the forces in the system are internal and no additional mechanical boundary conditions need to be applied to simulate stress-free boundaries of the tissue. 

We study two different fiber architectures: 1) all fibers being uniformly aligned in the $x$-direction and 2) fibers being organized in orthotropically stacked sheets, rotating about the $z$-axis through the thickness of the slab. For the second case, we assume that the fibers are parallel to the $x$-axis at the bottom (endocardium) and parallel to $y$-axis at the top surface (epicardium) with a total rotation angle of $90^{\circ}$ degrees.

\begin{table}[H]
	\caption{Parameters for the electrophysiology and MSM models. $\mathrm{n.u.}$ stands for no unit.}
	\label{Ta:electroparameters}
	\begin{ruledtabular}
		\begin{tabular}{l l l}
			 $U_h$ & $3$  & $\mathrm{n.u.}$\\ 
			 $U_v$ & $1$  & $\mathrm{n.u.}$\\ 
			 $U^*$ & $1.5415$  & $\mathrm{n.u.}$\\
			 $K_t$ & $0.0415$   & $\mathrm{n.u.}$\\
			 $\tau_U$ & $2.5$  & $\mathrm{ms}$\\ 
			 $\tau_v$ & $250$  & $\mathrm{ms}$\\
			 $Re$ & $0.8$ or $1.0$  & $\mathrm{n.u.}$\\
			 $M$ & $10$  & $\mathrm{n.u.}$\\
			 $D_{\parallel}$ & $1.1$  & $\mathrm{cm^2s^{-1}}$\\
			 $D_{\perp}$ & $0.22$ & $\mathrm{cm^2 s^{-1}}$\\
			 $C$ & 	$500$  & $\mathrm{cm s^{-1}}$\\
			 $D_e$& $11$  & $\mathrm{cm^2 s^{-1}}$\\
			 $a$ & $0.03$  &$\mathrm{cm}$\\
			 $dt$ & $3.27$  & $\mathrm{\mu s}$\\
		\end{tabular}
	\end{ruledtabular}
\end{table}

\section{Results}\label{sec:results}

\subsection{Two-dimensional isotropic and anisotropic tissues: emergence and properties of mechanical phase singularities}\label{sec:results2d}

\begin{figure}[]
\centering
\includegraphics[width=\columnwidth,angle=0,trim={0in 0in 0in 0in}]{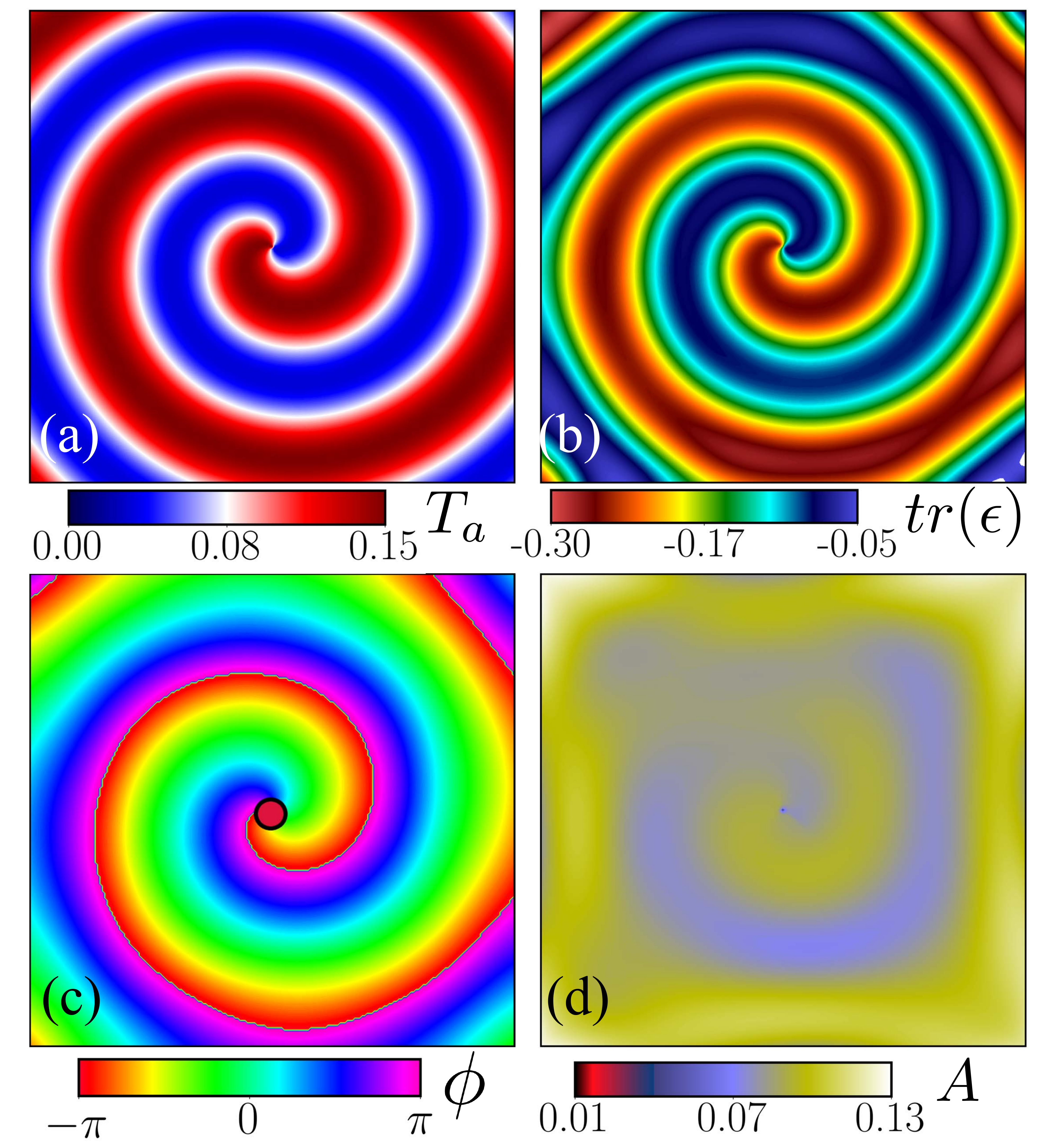}{}
\caption{A clockwise rotating spiral of active contraction and strain in a 2D isotropic elastic tissue. 
(a) Analytically derived active contraction field $T_a(x,y)$. 
(b) The corresponding strain field $\Tr(\epsilon)(x,y)$ is isotropic and closely follows the contraction field.
(c) The mechanical phase map $\phi(\epsilon)(x,y)$ reveals a mechanical phase singularity at the center of the rotor (red circle). 
(d) The complex amplitude $A(x,y)$ vanishes at the position of the mechanical phase singularity.}
\label{fig:2disospiral}
\end{figure}

\begin{figure}[]
\centering
\includegraphics[width=\columnwidth,angle=0,trim={0in 0in 0in 0in}]{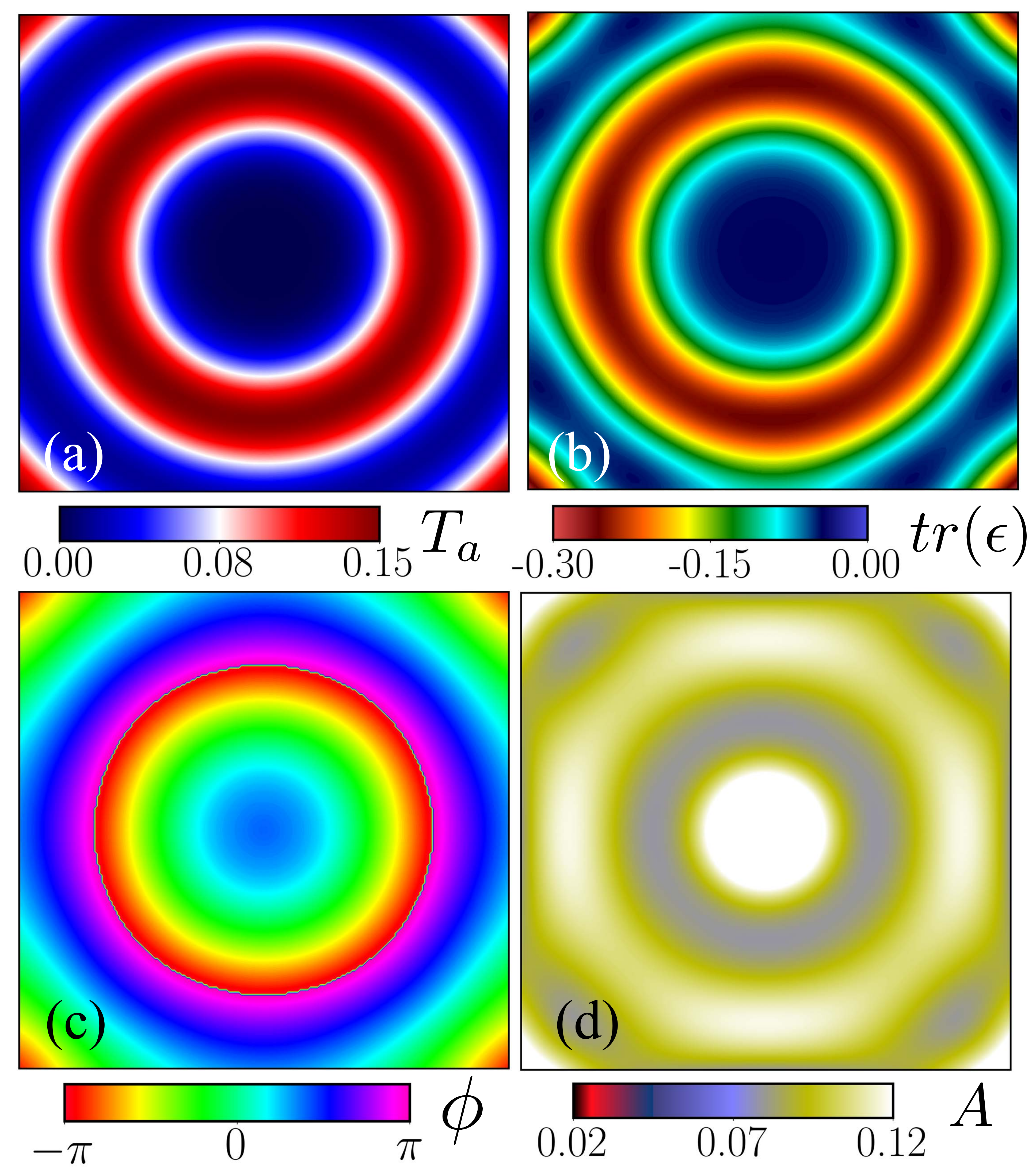}{}
\caption{Focal active contraction and strain pattern in a 2D isotropic elastic tissue.
(a) Circular ring-shaped focal pattern of active contraction $T_a (x,y)$.
(b) The corresponding strain field $\Tr(\epsilon)(x,y)$ is isotropic and closely follows the circular active contraction field.
(c) The mechanical phase map $\phi(\epsilon)(x,y)$ exhibits a focal pattern and does not exhibit a phase singularity at any point.
(d) The complex amplitude $A(x,y)$ does not vanish at any point.}
\label{fig:2disopace}
\end{figure}

\begin{figure}[]
\centering
\includegraphics[width=\columnwidth,angle=0,trim={0in 0.7in 0.3in 0.5in}]{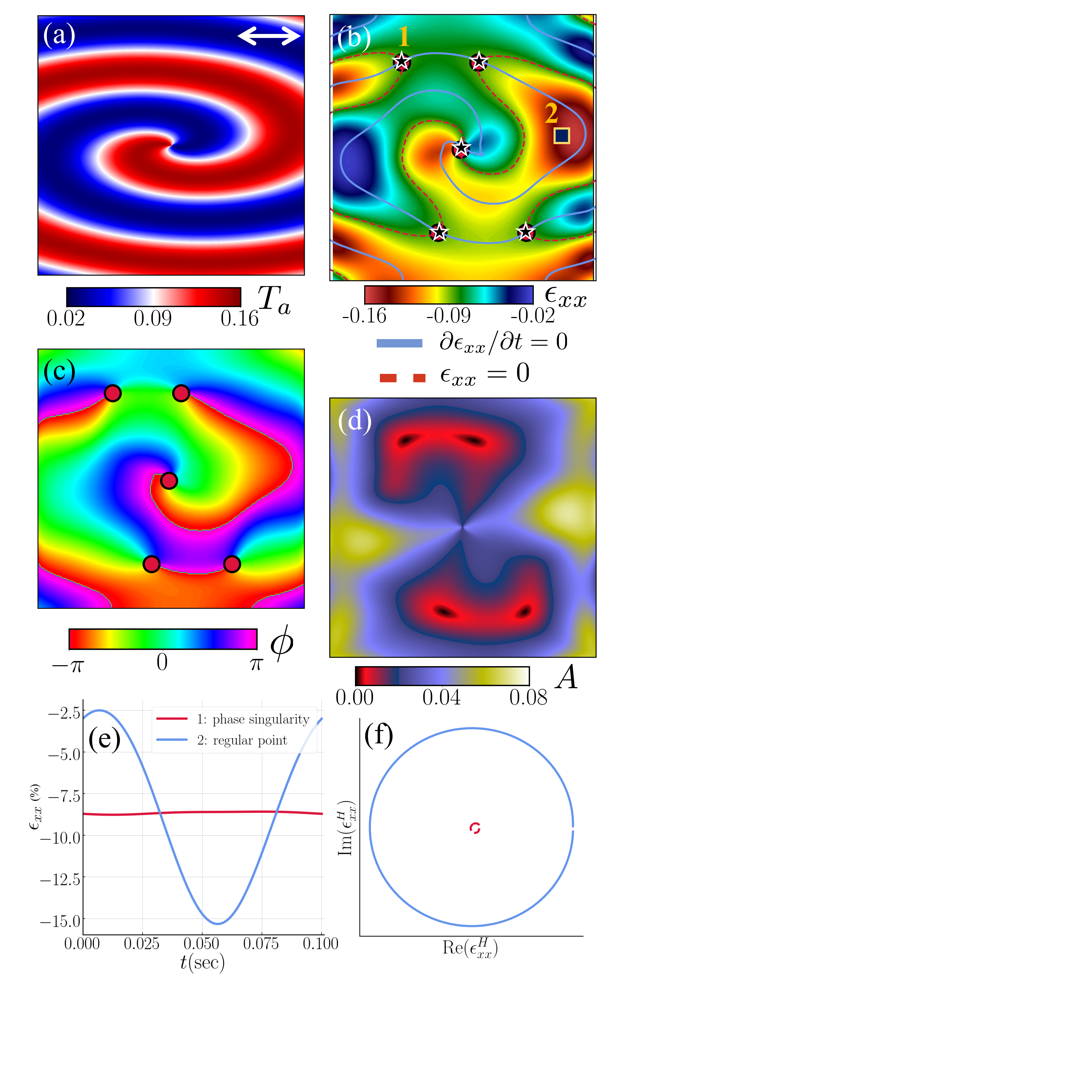}{}
\caption{\AMT{Strain spiral wave pattern in an anisotropic 2D medium with fibers aligned along the $x$-axis.
(a) Clockwise rotating spiral wave pattern of active contraction $T_a (x,y)$.
(b) Strain along the fibers $\epsilon_{xx}$ vanishes in the horizontally aligned arms. Red circles and black stars with white edge indicate the location of mechanical phase singularities calculated via the Hilbert transform and as the intersection of the $\epsilon_{xx}=\epsilon_c$ contour (thin red dashed lines) and $\partial_t\epsilon_{xx}=0$ contours (thin blue lines), respectively, where $\epsilon_c$ is the average of the minimum and maximum values of $\epsilon_{xx}$.
(c) Phase map of the strain $\phi ( \epsilon_{xx}(x,y))$ shows one co-localized mechanical phase singularity at the center of the rotor pattern in (a) and two additional pairs in the top and bottom horizontally aligned spiral arms. 
(d) The amplitude $A$ of the complex transformed signal vanishes in the vicinity to all mechanical singularities (black regions).
(e) Time-series of $\epsilon_{xx}$ for a points 1 and 2 in (b). 
(f) Phase diagram at points 1 and 2 from panel (e) derived via the Hilbert transform, where the horizontal axis is the original signal and the vertical axis is the signal shifted by $\pi/2$.}}
\label{fig:2danisospiral}
\end{figure}

\begin{figure}[]
\centering
\includegraphics[width=\columnwidth,angle=0,trim={0in 0in 0in 0in}]{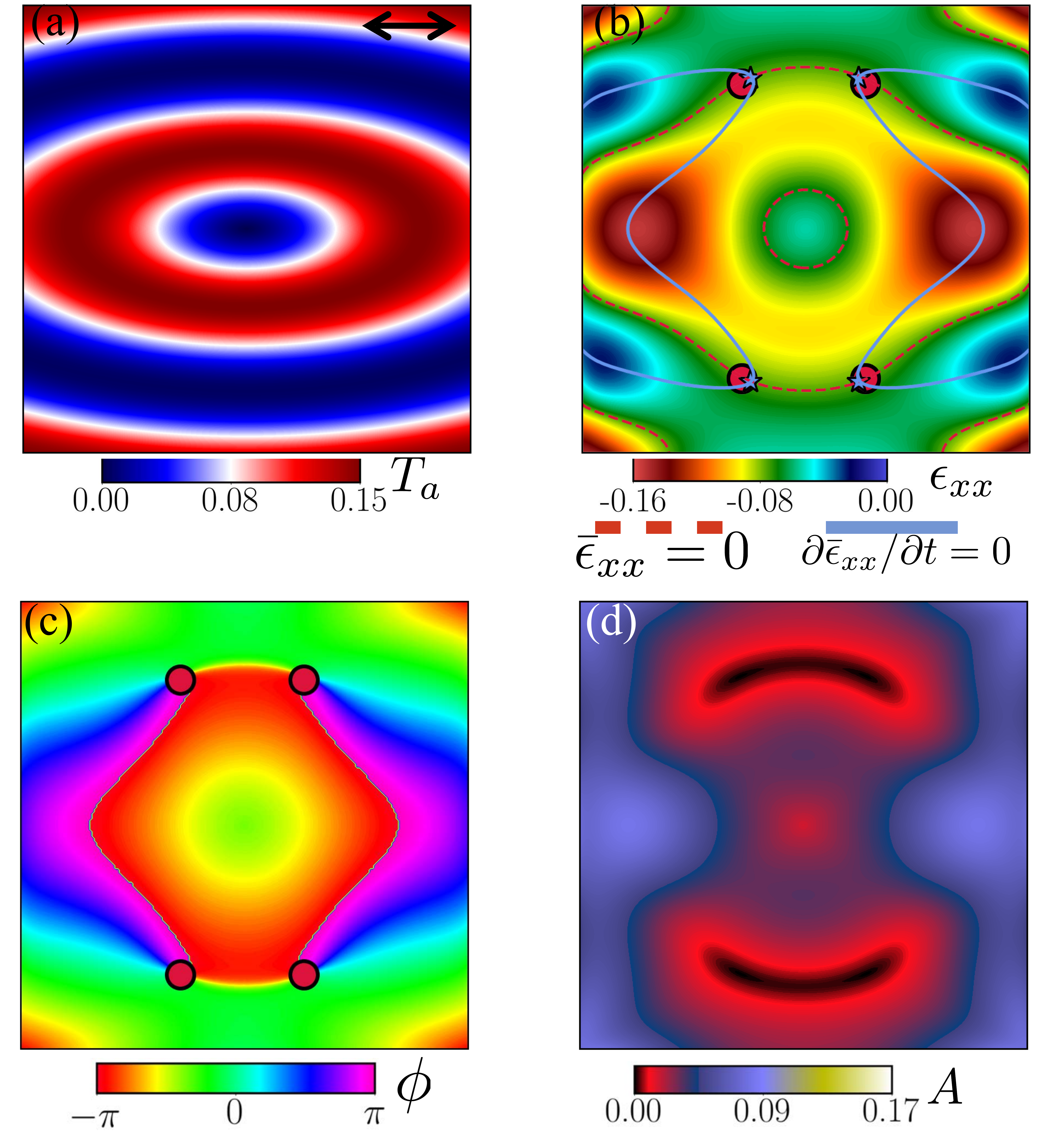}{}
\caption{
Focal wave in anisotropic medium with fibers aligned along $x$-axis.
(a) Active contraction field $T_a(x,y)$ of an elliptical ring-shaped focal pattern. 
(b) Strain $\epsilon_{xx}$ is stronger in the vertical than the horizontal arms due to fiber anisotropy. 
Red circles and blue stars indicate the location of phase singularities calculated by the Hilbert and contour-intersection method as in Fig.~\ref{fig:2danisospiral}, respectively.
(c) The phase map shows mechanical singularities emerge in the absence of electrical singularities. 
(d) The signal amplitude $A$ vanishes close to and around these singular points (black region).}
\label{fig:2dpace}
\end{figure}

\begin{figure}[]
\centering
\includegraphics[width=\columnwidth,angle=0,trim={0in 0in 0in 0in}]{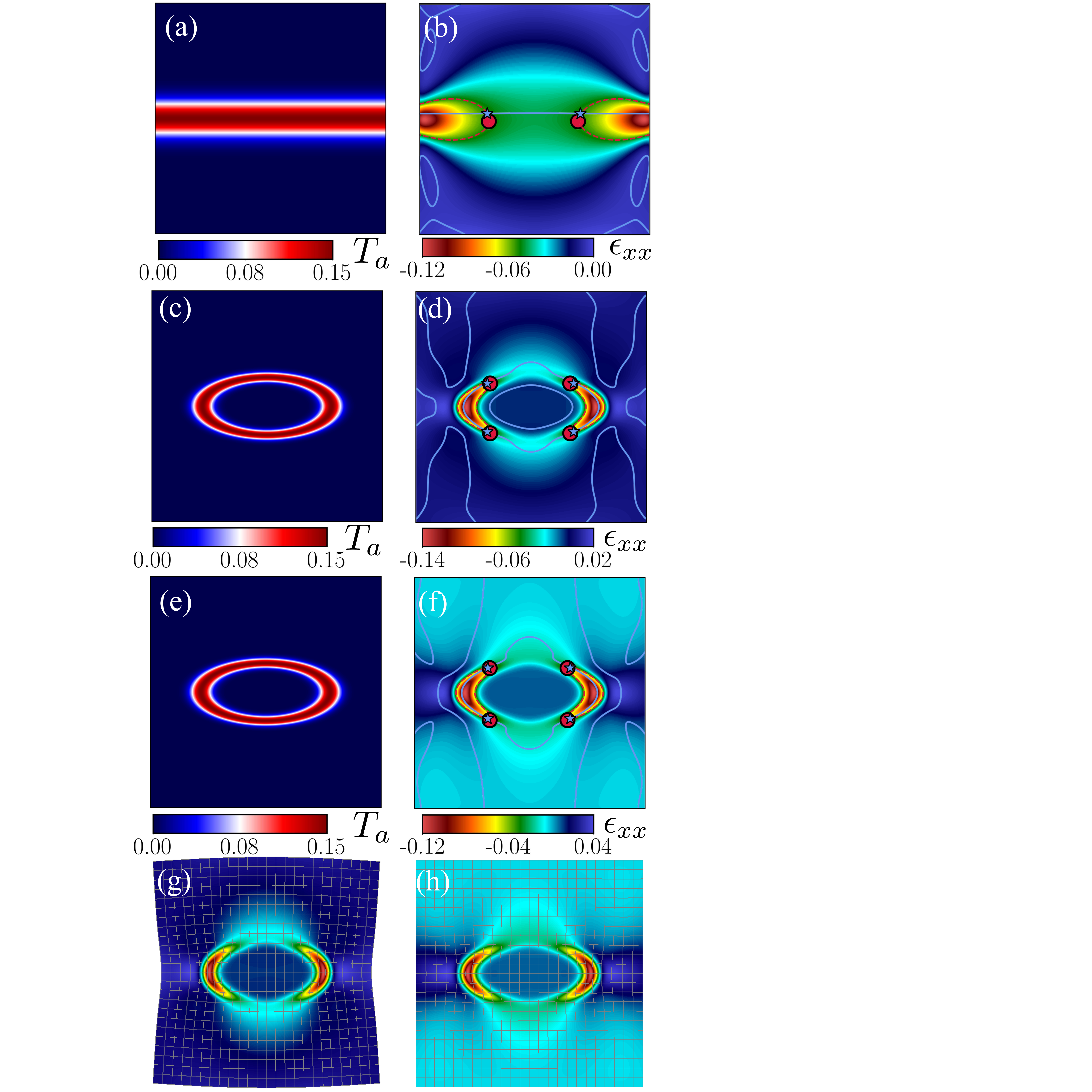}{}
\caption{Mechanism of formation of unpaired mechanical singularities in anisotropic tissue  
with fibers aligned along the $x$-axis. 
Active tension (a) and strain $\epsilon_{xx}$ (b) for a plane wave traveling in the $+y$-direction in a tissue with stress-free boundary conditions.
Active tension (c) and strain $\epsilon_{xx}$ (d) for a target wave expanding in a tissue with stress-free boundary conditions.
Active tension (e) and strain $\epsilon_{xx}$ (f) for a target wave expanding in a tissue with fixed boundary conditions ($u_x=u_y=0$ on all boundaries).
Red circles and blue stars indicate the location of phase singularities calculated by the Hilbert and contour-intersection method as in Figs.~\ref{fig:2danisospiral} and~\ref{fig:2dpace}, respectively. Panels (g) and (h) show the same strain patterns as (d) and (f), respectively, in deformed coordinates.}
\label{fig:planewave}
\end{figure}

We first study spiral-shaped and circular-shaped focal electromechanical wave patterns in 2D isotropic tissues, see Figs.~\ref{fig:2disospiral}-\ref{fig:planewave}.
\AMT{Before we begin to present the result, we want to emphasize that the study conducted in this paper is not limited to a particular strain signal such as $\Tr(\epsilon)$ or $\epsilon_{xx}$ or any particular direction, but it works for any signal that captures the mechanical response of the system. 
In this article we chose the strain along the fibers when there exists a fiber architecture (anisotropic case) or $\Tr(\epsilon)$ when there is not (isotropic case) to capture the mechanical response of the system.}
In the isotropic case, there is no particular fiber orientation ($f_i=1$ in~Eq.~\eqref{eq:eigen-strains}) and active tension occurs equally in all directions.
Moreover, instead of modeling the electrical spiral by using reaction-diffusion kinetics, we directly derived the shape of the active contraction field $T_a$ analytically using a closed-form expression presented in Eq.~\eqref{eq:spiralequation} in Appendix~\ref{app:spiral}.
Panels a) and b) in Fig.~\ref{fig:2disospiral} show the contraction field $T_a(x,y)$ and the resulting deformation of the tissue displayed as strain $\epsilon (x,y)$ in material coordinates, respectively.
Note that, because the contraction is isotropic, we chose the trace of the strain tensor $\Tr(\epsilon)(x,y) = \epsilon_{xx} + \epsilon_{yy}$ to visualize the strain as a scalar-valued field, where negative values correspond to contracted tissue.
In the isotropic case, it can generally be observed that the spiral shape emerging in the strain field matches the spiral shape in the active contraction field very well.
The high correlation between the scalar-valued contraction field and a scalar-valued representation of the strain furthermore manifests when comparing time-series obtained from individual tissue segments: over several rotations of the spiral, each local rise and decline in active contraction $T_a(t) \in [0,0.15]$ causes a corresponding shortening and elongation of the same tissue segment, which can be measured as a correspondingly oscillating strain signal $s(t) = \Tr (\epsilon) (t)$ which aligns with the time-course of the contraction variable $T_a (t)$ for each and every material segment throughout the tissue.
However, despite the strong correlation between contraction and resulting strain, the spiral shape in the strain field in (b) does not perfectly match the contraction spiral in (a), and scaling the fields would be insufficient to map one pattern onto another. Upon close inspection, the strain spiral pattern in Fig.~\ref{fig:2disospiral}(b) exhibits slight distortions and deviations, which can be observed best close to the tissue boundaries, c.f. Fig.~\ref{fig:2disopace}(b).
This behavior is to be expected because the overall strain field results from a superposition of several elastic phenomena which are all long-range effects interacting with each other over long distances throughout the entire medium: local contraction-induced deformations, such as contracted or elongated tissue regions pulling and pushing each other, and mechanical boundaries restricting the deformation. Note, that the strain fields are shown in Figs.~\ref{fig:2disospiral}-\ref{fig:planewave} are obtained at equilibrium and are fully relaxed.

Next, we use the Hilbert transform to assign a phase angle $\phi (t)$ and its complex amplitude $A(t)$ to each deformation state of each material segment throughout the medium yielding a phase map $\phi (x,y,t)$ and an amplitude map $A(x,y,t)$ as shown in panels (c) and (d) in Fig.~\ref{fig:2disospiral}, respectively.
The phase $\phi (t) $ and amplitude $A(t)$ are calculated individually at each point $(x,y)$ respectively:
\begin{align}
&\phi (t) =\arctan\left\{\frac{\Im(\tilde{s}(t))}{\Re(\tilde{s}(t))}\right\},\label{eq:phase}\\
&A (t) =\sqrt{\Im(\tilde{s}(t))^2+\Re(\tilde{s}(t))^2},\label{eq:amplitude}
\end{align}
where $\tilde{s}(t) = \mathcal{H}(s(t))$ is the Hilbert transform of an arbitrary signal $s(t)$, which is in our study either a mechanical signal derived from a deformation tensor or the time-course of the transmembrane potential. The phase angle $\phi (t)$ continuously increases over time from $-\pi$ to $\pi$ and uniquely represents the time-course of the action potential or a mechanical state, for instance, the deformation state within a cycle defined by two consecutive fully contracted states, see isoline for $\phi = - \pi$ in Fig.~\ref{fig:2disospiral}(c).
As stated above, with isotropic tissues, we chose the mechanical signal to be the temporal evolution of the rotationally invariant quantity $s(t) = \Tr (\epsilon)(t)$.
In anisotropic tissues, we adjusted the mechanical signal to comprise only mechanical strains occurring along the fiber direction.
For instance, with a uniform linearly transverse fiber orientation along the horizontal axis, we chose the signal to be $s(t) = \epsilon_{xx}(t)$ axis correspondingly.
In 3D anisotropic tissues, where the fiber direction rotates throughout the thickness of the bulk, we chose the signal to be the component of the Green-Lagrangian strain tensor corresponding to the local fiber axis, as later specified in section III.B.
Prior to computing the phase $\phi$ and amplitude $A$, we subtracted the temporal average of the mechanical signal in each point: 
\begin{align}
s(t) = s'(t) - \langle s' (t) \rangle_t 
\end{align}
taken over the same time interval that the Hilbert transform is computed, where $s'$ is the original mechanical signal with a finite baseline.
This procedure is used to determine the signal amplitude oscillating around its mean value at each point because time-averaged values of the signal can in general vary in space throughout the medium. 
Phase singularities can then be found in the phase maps $\phi (x,y)$ by computing the line integral \cite{Gray1998}:
\begin{align}
\oint \nabla \phi(\vec{r})ds=2\pi(p-n)
\label{eq:singintegral}
\end{align}
where the integral is taken over a closed path and $p$ and $n$ are the numbers of positively and negatively charged phase singularities inside the area enclosed by the path, respectively.

In the anisotropic case, we additionally compute mechanical phase singularities by a different method used previously to identify the {\it instantaneous} tip position of spiral waves of electrical activity \cite{Fenton:1998aa}. 
In that context, the spiral wave shape was defined as a contour of constant transmembrane voltage $U=U_c$, where $U_c$ is a constant chosen to lie in between the minimum and maximum values of $U$ corresponding to the resting voltage and the peak voltage following depolarization. 
This contour then separates resting and depolarized regions of the tissue and the spiral wave tip can be defined as the intersection of $U=U_c$ contour at time $t$ and $t+dt$, or equivalently as the intersection of the $U=U_c$ and $\partial_tU=0$ contours at time $t$. 
Similarly, we define the position of mechanical singularities as the intersection of the $\epsilon_{xx}=\epsilon_c$ and $\partial_t\epsilon_{xx}=0$ contours with the constant $\epsilon_c$ defined as the average of the minimum and maximum values of $\epsilon_{xx}$. 
With this definition, the $\epsilon_{xx}=\epsilon_c$ contour separates under- and over-strained regions of the tissue and mechanical singularities locate singular endpoints of this contour that can be interpreted as instantaneous deformation wave tips. 
The Hilbert transform and contour-intersection methods produce almost identical locations of mechanical singularities in most of the cases presented here where spatiotemporal patterns of voltage and mechanical activity are time-periodic. 
The contour intersection method is used here as additional validation of the Hilbert transform method as well as for physically interpreting the formation of unpaired mechanical phase singularities in terms of deformation wave breaks leading to the creation of deformation wave tips.

Figures~\ref{fig:2disospiral}(c--d) show the corresponding phase map $\phi$ and amplitude map $A$ of the strain pattern shown in Fig.~\ref{fig:2disospiral}(b). 
The phase map in Fig.~\ref{fig:2disospiral}(c) reveals a pinwheel pattern, which retains a similar spiral-like shape, as in Fig.~\ref{fig:2disospiral}(a--b) and exhibits spiral-shaped lines of equal phase merging at the center of the medium. 
Using Eq.~\eqref{eq:singintegral} one can detect a phase singular point (red dot) at the center of the medium, describing a topological defect point in the phase plane.
At this point, the elastic medium may be contracted or dilated, but the signal does not change with time, and consequently, the amplitude approaches zero, see Fig.~\ref{fig:2disospiral}(d).
Note that, due to the definition of amplitude (Eq.~\eqref{eq:amplitude}) in the Hilbert transformation, and the fact that we are working with the normalized response, amplitude being \textit{close to} zero means that there is \textit{almost} no difference between $\bar{s}$ and its $90^{\circ}$ shifted value (the difference between the real and imaginary parts of the Hilbert transformed signal). 
In other words, the signal remains \textit{almost} constant in time.
We will discuss this matter in more detail for the anisotropic spiral case.

What is clear from this simulation is that in the absence of fiber architecture (and consequently force anisotropy), the hydrostatic strain field $\Tr(\epsilon)$ follows the contraction field closely. 
As a result, the only mechanical phase singularity that we observe is due to the existing phase singularity of the prescribed active contraction force field.
Therefore, we can hypothesize that in tissue with isotropic fiber architecture and no singularity of contraction field, we must not observe any mechanical phase singularity.
To test this hypothesis, we apply an active contraction field of circular ring-shaped focal patterns to the tissue with isotropic fiber architecture.
Again, the active contraction field was derived analytically as presented in Eq.~\eqref{eq:elipsequation} in Appendix~\ref{app:spiral}.
The results of which are presented in Fig.~\ref{fig:2disopace}.
As one can see in Fig.~\ref{fig:2disopace}, the contraction field $T_a(x,y)$ in (a), the strain field $tr(\epsilon)(x,y)$ or trace of the strain tensor in (b), and the phase map of the strain $\phi ( \Tr(\epsilon))(x,y)$ in (c) all retain the circular shape, and there is no phase singularity in the phase map.
Moreover, the amplitude of the transformed signal remains relatively large everywhere inside the tissue.

In the cardiac muscle, the fiber architecture is highly anisotropic and organized in a complex, orthotropic pattern.
Furthermore, the contracting active force is exerted only along the local fiber direction.
As the next step, we study the effect of fiber anisotropy in the 2D tissue with a clockwise rotating spiral of active contraction.
With no loss of generality, we assume that the fibers are aligned uniformly along the $x$-axis, and therefore, the contractile force occurs in this direction as well.
The results for this case are presented in Fig.~\ref{fig:2danisospiral}.
Figure~\ref{fig:2danisospiral}(a) shows the active contraction field $T_a (x,y)$.
Note that the active contraction pattern is elongated along the horizontal $x$-axis, which reflects the underlying horizontal fiber anisotropy. 
The elongation occurs in real myocardium because the conduction is higher along the fiber direction, and therefore, electrical current propagates faster in this direction.
Since active contraction is slaved to electrical current, the active contraction will propagate faster and elongates along the same axis as well.
Correspondingly, in the anisotropic simulation, the contractile force is only exerted along the horizontal $x$-axis, which represents myocardial muscle tissue more accurately than the isotropic simulations shown in Figs.~\ref{fig:2disospiral} and~\ref{fig:2disopace}.
To identify mechanical phase singularities, while accounting for the effects caused by the fiber anisotropy, we analyzed the corresponding strain along the fiber orientation, which in this case is $\epsilon_{xx}$.
Due to the effects of force anisotropy along the horizontal axis, the strain pattern in Fig.~\ref{fig:2danisospiral}(b) is distorted compared to Fig.~\ref{fig:2disospiral}(b), and the generated strains are generally larger in the vertical arms.
In other words, while the contraction field is similar in both vertical and horizontal spiral arms in both the aniso- and isotropic case, the strain is significantly diminished in the horizontal parts of the spiral arms in the anisotropic case because the tissue cannot generate sufficient contractile force perpendicularly to the fiber orientation.
In comparison to the isotropic case, the phase map $\phi(x,y)$ in Fig.~\ref{fig:2danisospiral}(c) is more complex, and it does not retain a simple pinwheel shape.
The phase map reveals that, next to a mechanical phase singularity close to the core of the spiral pattern, additional mechanical phase singularities form at a distance from the central phase singularity in the horizontal spiral arms.
The central mechanical phase singularity co-localizes with the rotational core or tip of the electrical (or active contraction) spiral pattern shown in Fig.~\ref{fig:2danisospiral}(a), but the additional mechanical phase singularities do not co-localize with any phase singularities in the active contraction field (movies are available in SI).
In this example, see Fig.~\ref{fig:2danisospiral}(c), two pairs of additional mechanical phase singularities form in the vertical direction on each side of the core.
The two singularities in each pair exhibit opposite topological charges, but do not exhibit vorticity as the central stationary phase singularity does.
Overall, one notices an underlying pinwheel pattern, as in Fig.~\ref{fig:2disospiral}(b--c), but the pattern is topologically discontinuous in the space between each of the singularity pairs.
Furthermore, the amplitude A in Fig.~\ref{fig:2danisospiral}(d) vanishes in the vicinity of the mechanical phase singularities, independently of whether they are paired or unpaired with singularities of the contraction field.
However, in the surrounding neighborhood of paired and unpaired mechanical phase singularities, the spatial distribution of the amplitude $A$ differs significantly. 
Near paired singularities, the amplitude increases rapidly with distance from the singular point, while the amplitude vanishes at the center of the singular point in Fig.~\ref{fig:2danisospiral}(d)).
Near unpaired singularities, the amplitude remains small over a spatially extended region surrounding the singular point (black and red regions in Fig.~\ref{fig:2danisospiral}(d)).
As we show later, the property that the amplitude of oscillation grows rapidly or slowly away from paired or unpaired singularities, respectively, can be exploited to distinguish between those two different types of mechanical phase singularities. 

To better understand the nature of unpaired mechanical phase singularities, we examine the strain's temporal evolution (measured along $\epsilon_{xx}$, here aligned with the fiber direction) close to and further away from an unpaired singularity, see Fig.~\ref{fig:2danisospiral}(e).
One can observe that in the vicinity of an unpaired mechanical singularity (\eg point 1) the amplitude of $\epsilon_{xx}$ vanishes over the entire period, whereas further away at a non-singular site $\epsilon_{xx}$ exhibits large oscillations (\eg point 2), c.f. Fig.~\ref{fig:2danisospiral}(b).
The inset in Fig.~\ref{fig:2danisospiral}(e) shows how the corresponding trajectories of these two time-series look in a two-dimensional complex phase space with the real and imaginary parts of the signal on the $x$- and $y$-axis respectively: the non-vanishing time-series of the strain with finite amplitudes corresponds to a circular trajectory, whereas the strain with vanishing amplitude reduces to a point in phase space, which prohibits defining a phase angle $\phi (\epsilon_{xx})$.
This behavior is rather unexpected for an unpaired singularity as shown in point 1 in Fig.~\ref{fig:2danisospiral}(e). 
As we discuss next for the simpler case of a focal excitation, this property is the result of the combined effects of fiber anisotropy and the non-locality of elastic interactions that can cause $\epsilon_{xx}$ to remain constant in time. 

The strain field resulting from a focal excitation wave in an anisotropic medium is shown in Fig.~\ref{fig:2dpace}.
The contraction field $T_a (x,y)$ and the strain field measured along the fiber axis $\epsilon_{xx} (x,y)$ are shown in (a) and (b), respectively.
Note that the thickness of the focal contraction wave is not uniform due to the differences in conduction speed of the excitation wave along the horizontal and vertical directions, respectively, similarly as in Fig.~\ref{fig:2danisospiral}.
As a result, the wave is thicker in regions in which it propagates in parallel to the fiber axis than in regions in which it propagates perpendicularly to that axis. 
As can be seen in Fig.~\ref{fig:2dpace}(b), the strain in the vertically traveling parts of the elliptical wave is diminished.
Importantly, panel (c) shows that the elliptical focal contraction wave produces four unpaired mechanical phase singularities in the tissue.
These four phase singularities emerge in pairs and are aligned symmetrically around the focal center of the wave and travel along a confined closed loop (movies are available in SI).
Lastly, panel (d) shows that the amplitude $A$ of the transformed signal vanishes in the neighborhood of these phase singularities, as was the case with the anisotropic spiral wave pattern shown in Fig.~\ref{fig:2danisospiral}.

To gain further insight into the mechanism of the formation of unpaired singularities, we paced an anisotropic tissue with a periodic wave train of electrical plane waves propagating from bottom to top along the $y$-direction while the fibers were aligned uniformly along the $x$-axis perpendicularly to the wave. 
A snapshot of this simulation is shown in Figs.~\ref{fig:planewave}(a-b). While the active tension is spatially uniform parallel to the fiber axis in the depolarized contracting region of tissue, see Fig.~\ref{fig:planewave}(a), the resulting deformation wave measured by 
$\epsilon_{xx}$ is ``broken'' in the middle region, see Fig.~\ref{fig:planewave}(b). 
This wave break gives rise to the formation of two mechanical singularities located at the tips of the $\epsilon_{xx}=\epsilon_c$ contours (thin red dashed lines) separating under-strained and over-strained regions of tissue shown in green and red in Fig.~\ref{fig:planewave}(b), respectively. 
This wave break occurs because the active force in the central region of tissue is contracting against large regions of passive surrounding tissue, thereby producing small strains, while less constrained regions closer to the stress-free boundaries of the tissue produce a larger strain. 
In contrast, with fixed displacement boundary conditions on all edges of the tissue ($u_x=u_y=0$), $\epsilon_{xx}=0$ in the whole tissue is the trivial solution of linear elasticity under the same plane wave of active tension that does not produce any strain inhomogeneities. 

However, unpaired mechanical singularities form independently of the particular mechanical boundary conditions with more complex wave patterns, such as the spiral and focal wave patterns. 
Figs.~\ref{fig:planewave}(c-f) show how unpaired mechanical phase singularities emerge with a focal wave both with stress-free boundary conditions (d) and fixed displacement $u_x=u_y=0$ boundary conditions (f). 
With stress-free boundary conditions, the tissue can contract freely without being restricted by the boundaries. All simulations in Figs.~\ref{fig:2disospiral}-\ref{fig:2dpace} were obtained with stress-free boundary conditions.
With fixed displacement boundary conditions the boundaries are fixed in space and do accordingly not give in to contractions occurring within the tissue. Panels (d) and (f) show that unpaired mechanical singularities emerge in both cases even though the strain fields are significantly different with the two mechanical boundary conditions.
All maps presented in this study so far were shown in undeformed coordinates. 
To illustrate the effect of the boundary conditions on the deformation, Figs.~\ref{fig:planewave}(g) and (h) show the corresponding deformation caused by the focal wave in (c) in deformed coordinates. With stress-free boundary conditions, the tissue's deformations are clearly visible, whereas with fixed boundary conditions the tissue's outer boundaries appear static while further inside the tissue deforms.
The results show that, independently of the choice of boundary conditions, four unpaired mechanical singularities are formed by the breakup of the two elongated thin regions of the elliptical target wave propagating upwards and downwards. Breakup occurs because regions of the elliptical target wave propagating along the fiber axis are thicker and hence exert a larger net contractile force that overstrains those regions. 
In contrast, regions propagating perpendicular to this axis are thinner, thereby exerting a smaller net contractile force. Those regions are also more elongated, thereby contracting against a larger region of passive tissue. Both factors contribute to causing those thinner elongated regions to be understrained. A similar mechanism is seen to underlie the formation of unpaired mechanical singularities in the case of reentrant spiral waves, see Fig.~\ref{fig:2danisospiral}, that, like elliptical target waves, exhibit both thicker and thinner regions propagating parallel and perpendicular to the fiber axis.

\begin{figure}[]
\centering
\includegraphics[width=\columnwidth,angle=0,trim={0in 0in 0in 0in}]{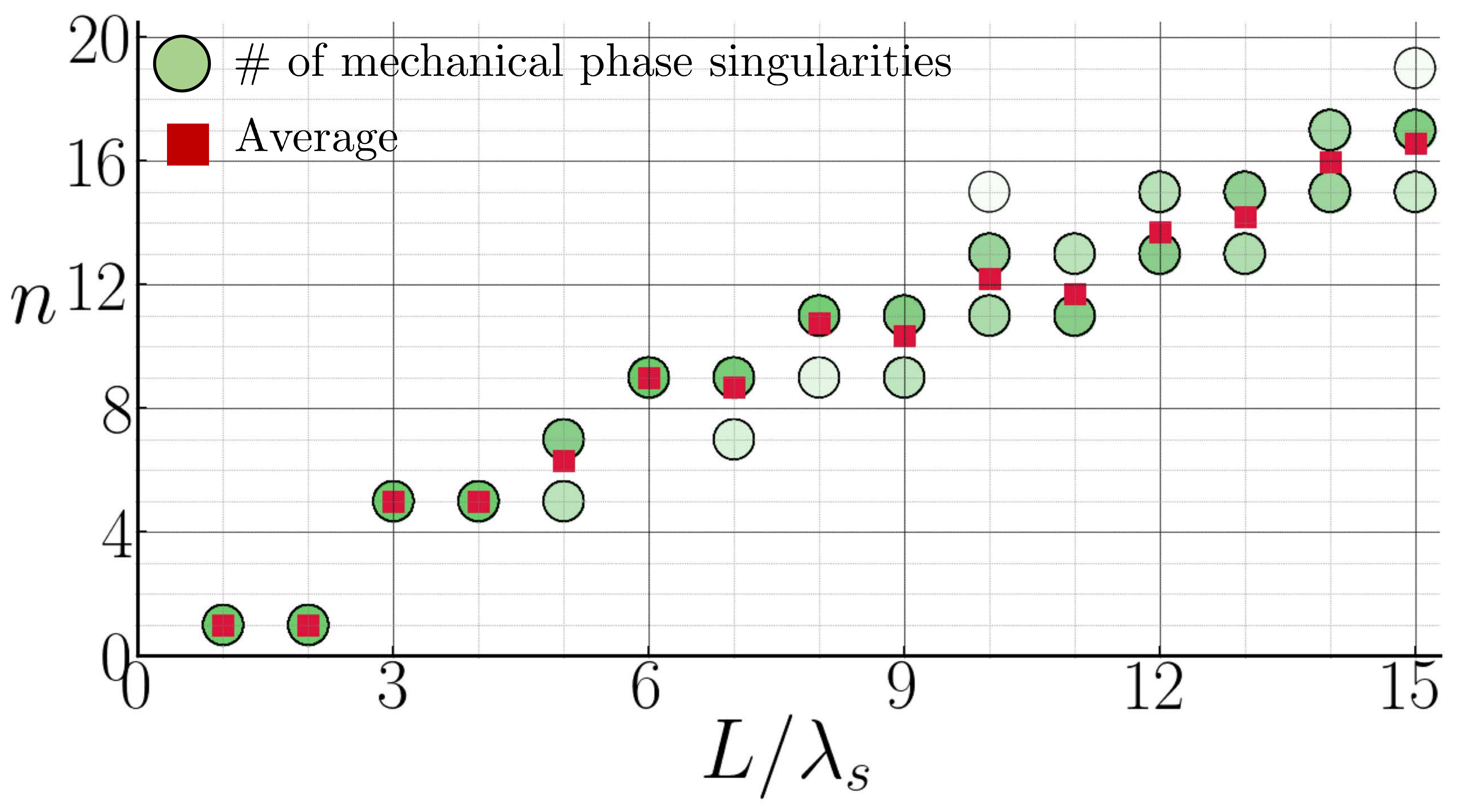}{}
\caption{Number of mechanical phase singularities for a clockwise rotating spiral in a heart muscle with anisotropic fiber architecture (fibers are along x-axis) as a function of tissue size and tissue size to spiral wavelength ratio $L/\lambda_s$.
Different vertical circles show the different number of phase singularities observed in one period.
The intensity of the color in each circle exhibits the frequency of occurrence of that number of phase singularities.
For instance, for ($L/\lambda_s=15$) during one period 15, 17, and 19 phase singularities have been observed.
The color density of the circles shows that most of the time there are 17 phase singularities in the domain.
The red squares show the average number of singularities during one period.}
\label{fig:singunum}
\end{figure}

We further investigated how the number of unpaired mechanical phase singularities increases with tissue size.
Figure~\ref{fig:singunum} shows the number of mechanical phase singularities per period as a function of ratio $L/\lambda_s$ of tissue size and spiral wavelength for a single spiral wave rotating in a 2D medium with anisotropic fiber architecture.
The results show an increase in the number of mechanical phase singularities with system size.
For small values of $L/\lambda_s$ ($L/\lambda_s=1,2$) the only mechanical phase singularity appears at the core of the spiral, which is associated with the co-localized singularity at the tip of the spiral.
As $L/\lambda_s=3$ the number of singularities jumps to 5 which accounts for the formation of two additional pairs of non co-localized singularities.
This case ($L/\lambda_s=3$) is the same as the one presented in Fig.~\ref{fig:2danisospiral}.
As $L/\lambda_s$ increases further, the mechanical state of the medium becomes more complex and pairs of mechanical phase singularities are created and annihilated during each period, thereby causing the number of unpaired singularities to fluctuate during one period. However, the average number of phase singularities increases roughly linearly with the number of turns of the spiral wave that increases proportionally to $L/\lambda_s$. This behavior is expected based on our physical interpretation of the formation of unpaired singularities based on deformation wave breaks. 

In summary, our 2D simulations demonstrate that mechanical phase singularities co-localize with the tip of electrical spiral waves and characterize the center of an electromechanical rotor.
In addition, unpaired mechanical phase singularities can form with both reentrant and focal waves, which then characterize the tip location of broken deformation waves.
 Wave breakup occurs due to the creation of overstrained and understrained regions of tissue that form preferentially along thicker and thinner regions of waves propagating parallel and perpendicular to the fiber axis, respectively. Since those waves can emanate from a focal source of excitation or a spiral wave center, the formation of unpaired mechanical singularity does not necessitate an electrical singularity to be present, thereby leading to a dissociation of a strict paired organization of electrical and mechanical phase singularities. In addition, mechanical boundary conditions can influence the formation of unpaired mechanical singularities, \eg by facilitating their formation near stress-free surfaces during plane wave propagation, but unpaired singularities form independently of the boundary conditions in the case of reentrant or focal excitation waves. 

\subsection{Three-dimensional Dynamics}\label{sec:results3d}
\begin{figure}[]
\centering
\includegraphics[width=\columnwidth,angle=0,trim={0in 0in 0in 0in}]{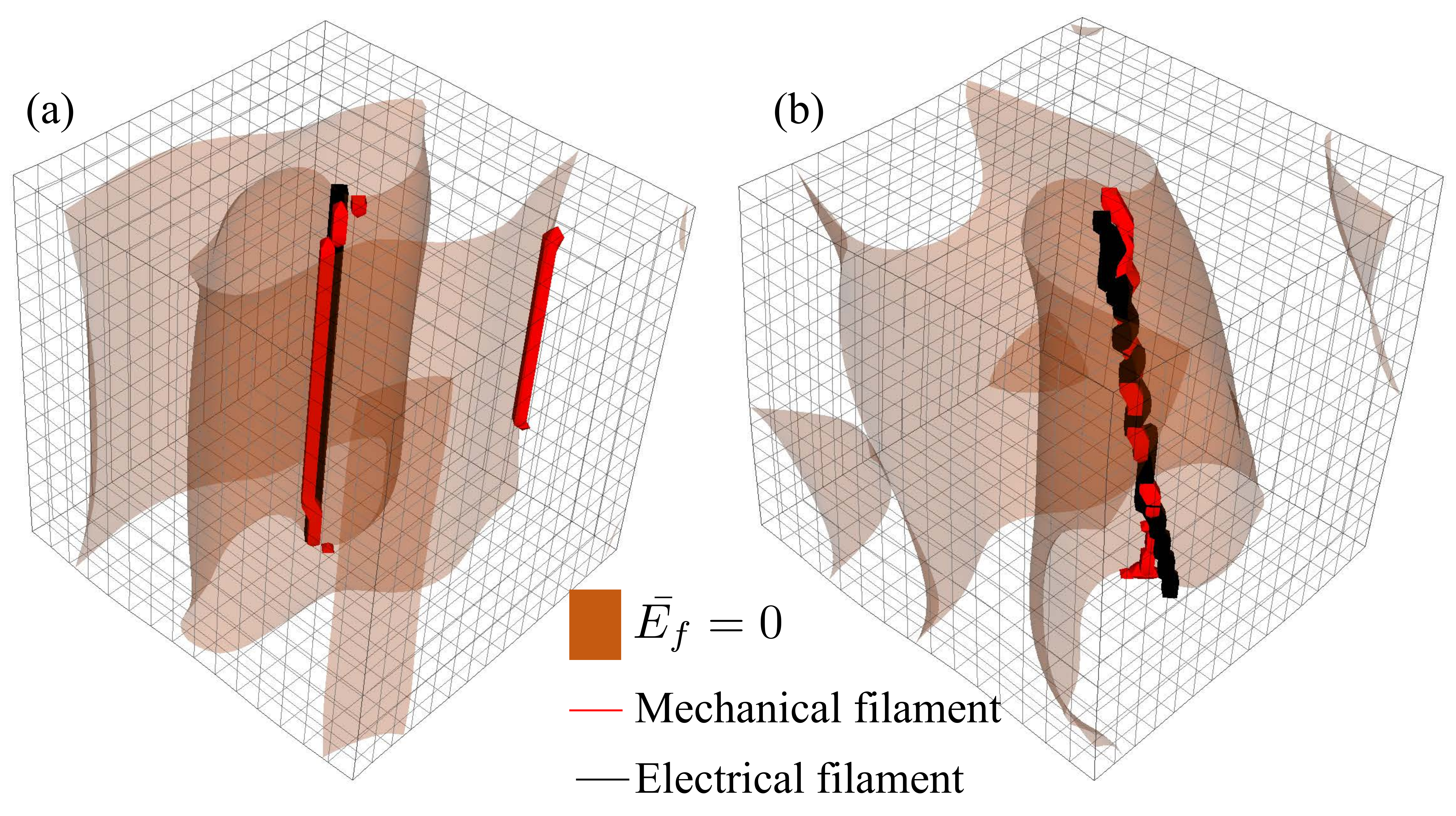}{}
\caption{
Electromechanical phase singularities in the presence of a single scroll wave in a 3D tissue ($1.2\times1.2\times1.2~\mathrm{cm}$).
(a) All fibers are uniformly aligned along the $x$-axis. 
(b) Fibers orthotropically stacked and rotating by $90^{\circ}$, parallel to $x$-axis at the bottom and parallel to $y$-axis at the top surface, respectively.
Red lines represent lines of mechanical phase singularity or mechanical filaments and the black line at the center represents the electrical filament or scroll wave core.
The iso-surface is presenting $\bar{E}_f=\bar{E}_{xx}=0$ where $\bar{E}_f$ is the normalized strain along the fiber that is used to calculate the phase.}
\label{fig:spiral3d_small}
\end{figure}

Phase singular points in 2D correspond to lines of phase singularity in 3D. 
Vortex dynamics in three-dimensional excitable media have frequently been characterized using lines of phase singularity, where the lines or vortex filaments indicate the rotational core regions of three-dimensional scroll waves \cite{Fenton:1998aa}.
Figure~\ref{fig:spiral3d_small} depicts electromechanical phase singularities in a three-dimensional bulk tissue ($1.2\times1.2\times1.2$~cm$^3$) with (a) uniformly horizontally aligned fibers or linearly transverse fiber architecture and (b) fibers being rotated by $90^{\circ}$ along the z-axis, where at the bottom face of the bulk they are aligned along the $x$-axis and at the top surface they are aligned along the $y$-axis. 
The tissue is deformed by a single electrical scroll wave.
The electrical scroll wave, which was simulated using the reaction-diffusion model presented in Sec.~\ref{sec:karma94}, can be described by a straight electrical vortex filament (black) being aligned vertically along the short axis in the center of the bulk in between its upper and lower surfaces. 
The electrical vortex filament does not bend or twist, and accordingly, the scroll wave remains stable.
There are co-localized mechanical phase singularities (red filaments) for both cases shown in (a) and (b).
However, in the case of no fiber rotation, one extra ''unpaired'' mechanical phase singularity in the form of a filament attached to the boundary can be observed.

{}

\begin{figure}[]
\centering
\includegraphics[width=\columnwidth,angle=0,trim={0in 0in 0in 0in}]{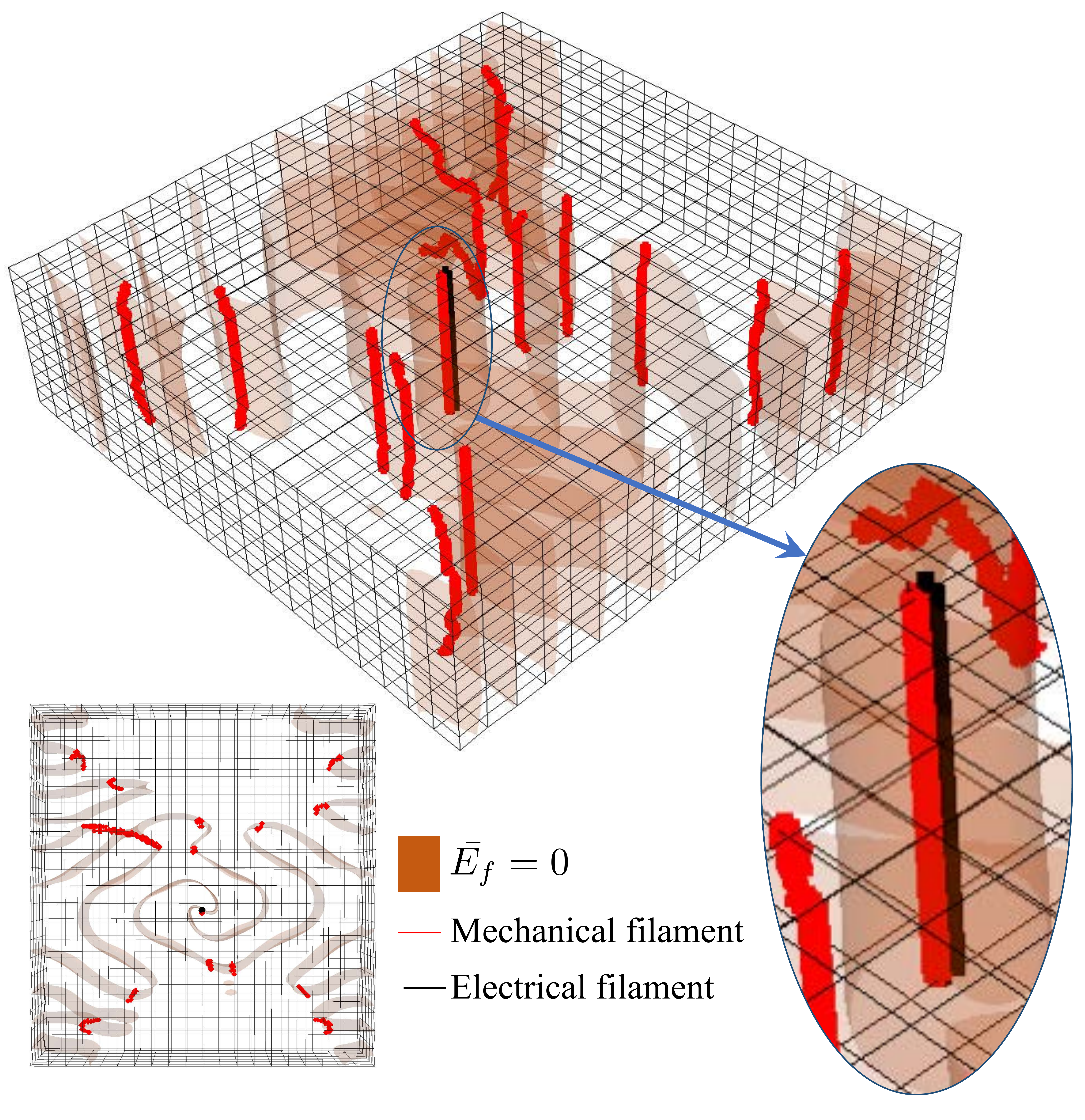}{}
\caption{Electromechanical phase singularities in the presence of a single scroll wave with short wavelength in a 3D tissue ($4.5\times4.5\times1.2$cm, inset figure is the top view) with all fibers being aligned along the x-axis.
Red lines represent lines of mechanical phase singularity or mechanical filaments and the black line at the center represents the electrical vortex filament or scroll wave core.
The iso-surface is presenting $\bar{E}_f=\bar{E}_{xx}=0$, where $\bar{E}_f$ is the normalized strain along the fiber that is used to calculate the phase.}
\label{fig:spiral3d}
\end{figure}

In these simulations, the ratio between the size of the tissue and the spiral wavelength falls in the left part of the graph shown in Fig.~\ref{fig:singunum}.
Therefore, there is only one pair of co-localized electrical and mechanical phase singularities describing the scroll wave core, if we ignore the one mechanical filament attached to the boundary.
To study the effect of tissue size and wavelength on the spatial organization of mechanical filaments, we carried out a simulation with the same parameters as in Fig.~\ref{fig:spiral3d_small}, but in a larger tissue with size $4.5\times4.5\times1.2 \, \mathrm{cm}$, see Fig.~\ref{fig:spiral3d}.
The filaments shown in Fig.~\ref{fig:spiral3d} for a single scroll wave were obtained with no fiber rotation as in Fig.~\ref{fig:spiral3d_small}(a).
A co-localized pair of an electrical and a mechanical filament mark the core region of the scroll wave at the center of the medium.
Additionally, one can see that there are several unpaired mechanical phase singularities outside of the core region in analogy to the 2D anisotropic case shown in Fig.~\ref{fig:2danisospiral}.

{}

\begin{figure}[]
\centering
\includegraphics[width=\columnwidth,angle=0,trim={0in 0in 0in 0in}]{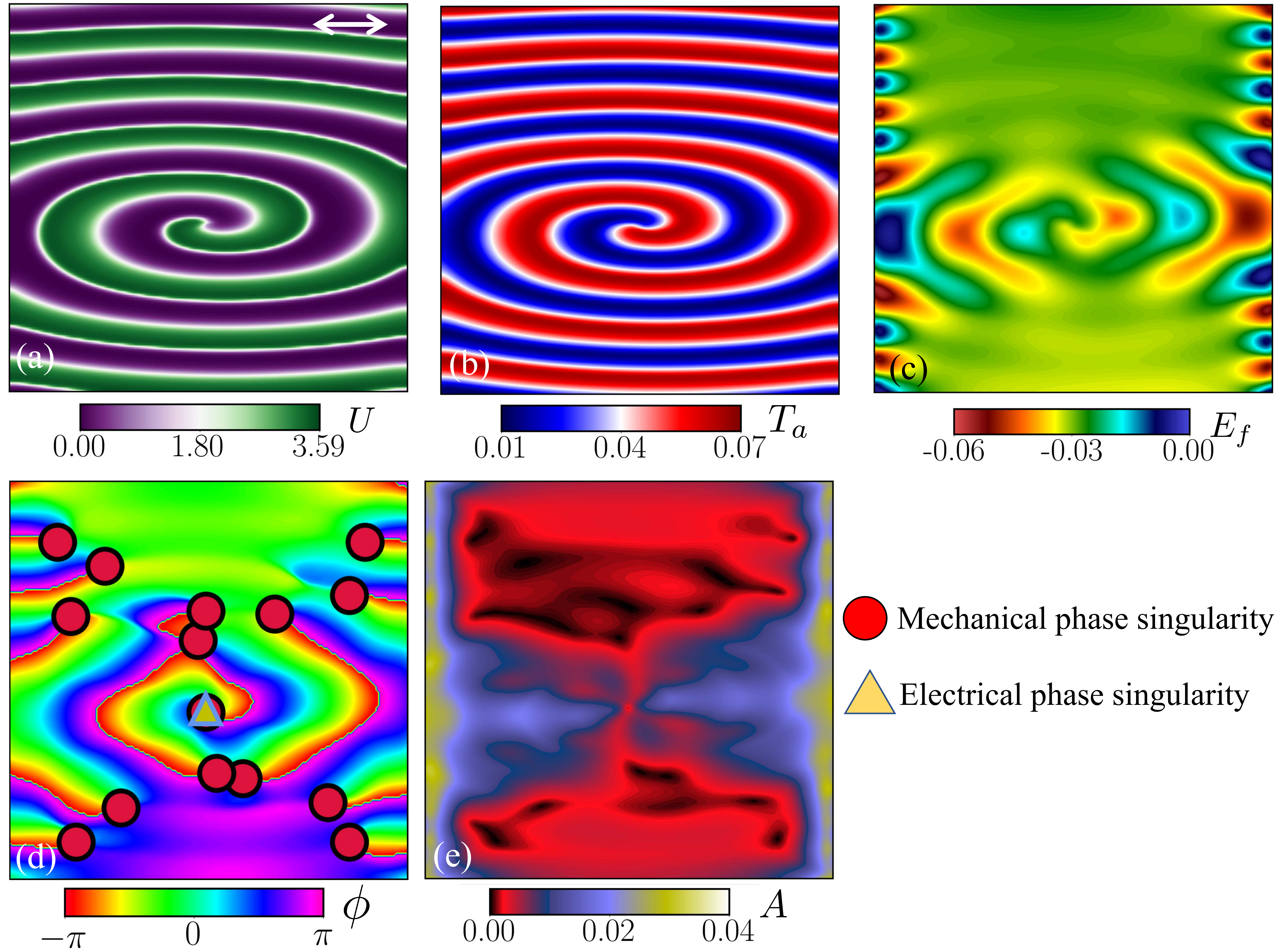}{}
\caption{Electromechanical organization of filament-like phase singularities on the surface of a 3D slab with single scroll wave with small wavelength and fibers in $x$-direction.
(a) Transmembrane voltage in dimensionless units. (b) Contraction field enslaved to voltage. (c) Strain depicted as largest compressive eigenstrain $E_{xx}$.
(d) Phase map $\phi(E)$ from the Hilbert transform of the normalized strain signal.
(e) Amplitude map $A(E)$ from Hilbert transform of the normalized strain signal. Amplitude vanishes in the neighbourhood of phase singular points.}
\label{fig:ionicspiralfiberx}
\end{figure}

The excitation, contraction, and strain fields together with the phase, singularities, and amplitude on the top surface of the tissue are illustrated in Fig.~\ref{fig:ionicspiralfiberx}. Panel (a) shows the electrical excitation $U$ (dimensionless normalized units n.u.) on the top surface of the ventricular tissue slab intersected by the vortex filament.
Due to the orientation of the scroll wave within the bulk, the near-surface electrical pattern corresponds to a clockwise rotating spiral wave pattern, and the tip of the spiral coincides with the intersection point of the vortex filament with the surface.
In addition, the electrical wave propagates faster along the $x$-axis parallel to the fibers (white arrow). 
This leads to a stretched spiral wave pattern in both electrical excitation $U$, as shown in (a), and active contraction $T_a$, as shown in (b).
To derive the strain field in (c) we computed the Green-Lagrangian strain tensor $\mathbf{E}$ as:
\begin{align}
\mathbf{E}:=\frac{1}{2}\left(\mathbf{F}^T\mathbf{F}-\mathbf{I}\right)
\end{align}
where $\mathbf{F}=[F_{ij}]=[\partial u_i/\partial x_j]$ is the deformation gradient tensor, and $\mathbf{I}$ is the identity. 
Unlike in 2D, the 3D simulations exhibit large strains. 
The strain pattern was derived by computing the strain along the fiber, here denoted as $E_{f}$, which in this case coincides with $E_{xx}$, similarly as shown in Fig.~\ref{fig:2danisospiral}(b). 
Note that the strain along the fiber \ie $E_{f}$, is a simple transformation of the strain tensor where the $x-$axis rotates to match the fiber direction.
However, because the wavelength of the spiral is small in comparison to the medium size, the spiral is curled up more, and the anisotropic strain pattern can be observed not only in the near-field of the spiral wave core but for multiple wavelengths.
The phase and amplitude maps presented in (d) and (e), respectively, show the existence of one central paired electromechanical singularity and a large number of mechanical phase singularities outside the spiral wave core that form because of the ratio of the tissue size to wavelength is sufficiently large in this simulation (similarly to Fig.~\ref{fig:singunum} in 2D). 

The spatial organization of unpaired singularities away from the central paired electromechanical singularity in Fig.~\ref{fig:ionicspiralfiberx} can be interpreted using the insights from Fig.~\ref{fig:planewave} that highlighted the roles of 2D wave patterns (plane waves or focal excitations) and mechanical boundary conditions in the formation of unpaired singularities. The unpaired mechanical singularities forming far from the spiral core form preferentially near the stress-free surfaces of the tissue due to the breakup of nearly plane waves of deformation by a mechanism directly analogous to Fig.~\ref{fig:planewave}(a-b), while mechanical singularities closer to the spiral core region form by a mechanism analogous to the one illustrated in Fig.~\ref{fig:planewave}(c-d) for a focal excitation, which also pertains to reentrant waves. Furthermore, the migration of unpaired mechanical singularities closer to stress-free surfaces with distance from the spiral core in Fig.~\ref{fig:ionicspiralfiberx}(d) can be interpreted as a transition between the near-core formation mechanism, which is insensitive to mechanical boundary conditions, and the far-core plane-wave breakup mechanism where stress-free surfaces favor the formation of unpaired singularities.

\begin{figure}[]
\centering
\includegraphics[width=\columnwidth,angle=0,trim={0in 0in 0in 0in}]{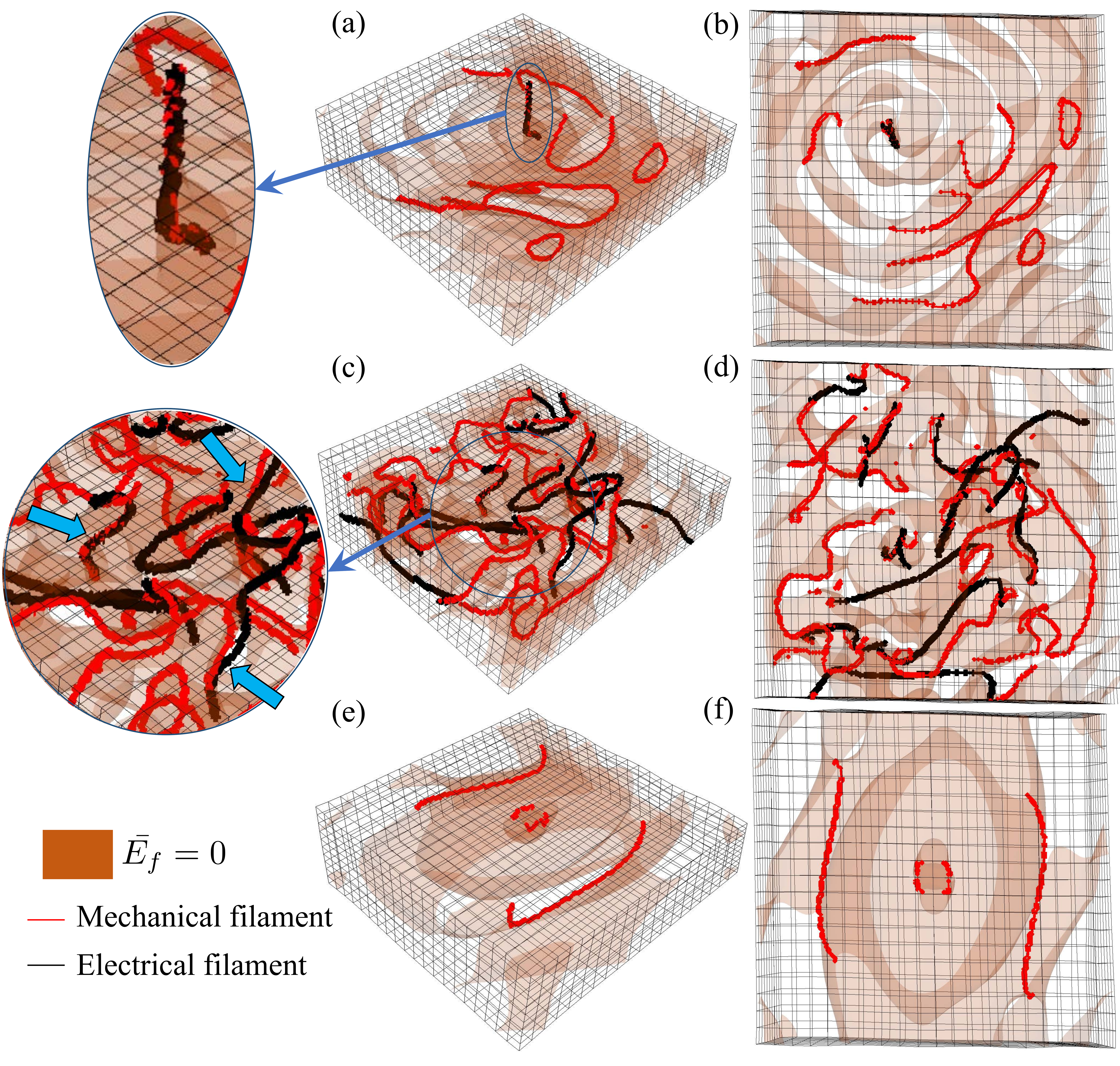}{}
\caption{Organization of electromechanical filaments in the presence of fiber rotation: 
(a-b) Single electrical scroll wave.
(c-d) Composition of multiple scroll waves due to electrical wave break. (e-f) Focal wave pattern after application of a pulse at the center of the top surface.
The isosurface is presenting $\bar{E}_f=\bar{E}_{xx}=0$ where $\bar{E}_f$ is the normalized strain along the fiber.
Insets show co-localized electromechanical filaments. In the case that electrical waves remain stable one can see a clear co-localization between mechanical and electrical filaments. 
However, as scroll waves break, a more complex mechanism can be observed, but many of electrical filaments are still paired with a mechanical one.}
\label{fig:mechbreakup}
\end{figure}

Finally, Fig.~\ref{fig:mechbreakup} depicts electromechanical vortex filament dynamics for the more realistic case that fibers are organized orthotropically and the fiber direction rotates along the $z$-axis through the thickness of the ventricular wall. 
In panels (a-f), with no loss of generality and for simpler visualizations, we show the normalized strain along the fiber $\bar{E}_f$. This measure was consequently used for further analysis using the Hilbert transform. 
At the bottom of the bulk at $z=0$, the fibers are aligned uniformly along the $x$-axis.
At the top, they are aligned along the $y$-axis.
This results in a $90^{\circ}$ rotation over the thickness of the bulk of $1.2\,\mathrm{cm}$.
Panels (a--b) show the electromechanical filaments being produced by one clockwise rotating electrical scroll wave pattern.
The comparison of Fig.~\ref{fig:mechbreakup}(a-b) and Fig.~\ref{fig:spiral3d} shows that the fiber rotation causes a more complex structure with bent filaments, but a qualitatively similar organization with a central paired electromechanical singularity (black and red) surrounded by additional unpaired mechanical singularities (red). 

Next, panels (c-d) in Fig.~\ref{fig:mechbreakup} show the filament dynamics in the regime of electrical wave turbulence corresponding to steeper action potential duration restitution slope ($Re=1$ instead of $0.8$, see Eq.~\eqref{eq:restitution}). 
In this regime, scroll wave breakup leads to a multiplication of electrical filaments. 
Interestingly, the results in Figs.~\ref{fig:mechbreakup}(c-d) show that the electrical (black) and mechanical filaments (red) are not as well co-localized in the scroll wave breakup regime compared to the single stable scroll wave regime of Figs.~\ref{fig:mechbreakup}(a-b). However, paired electromechanical singularities can still be distinguished from unpaired singularities that are not co-localized with any electrical filaments. 
Lastly, we show in Figs.~\ref{fig:mechbreakup}(e-f) the results obtained with a 3D focal wave after pacing the bulk at the center of its top surface. 
The results confirm that unpaired mechanical singularities can exist without the presence of electrical singularities as a direct extension of the 2D results shown in Fig~\ref{fig:2dpace}.

\begin{figure}[]
\centering
\includegraphics[width=\columnwidth,angle=0,trim={0in 0in 0in 0in}]{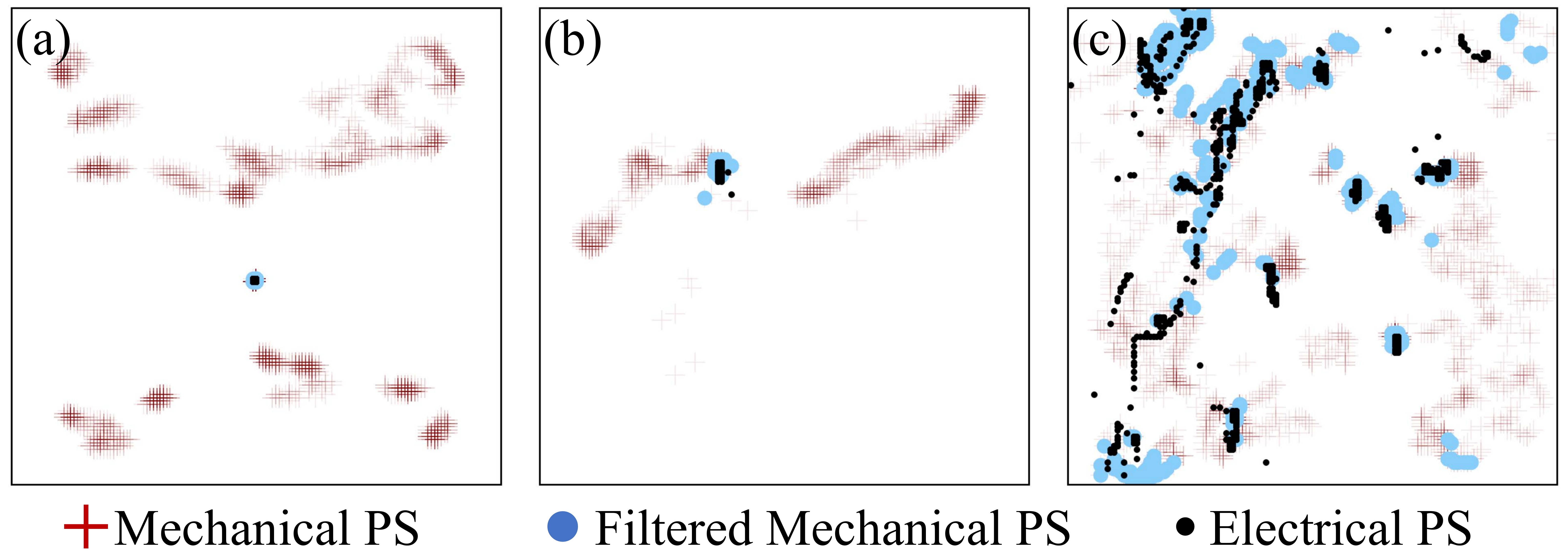}{}
\caption{
Automated separation of paired (blue dots) from unpaired mechanical singularities (red crosses). Paired mechanical singularities co-localize with electrical singularities (black dots).
The separation was performed by analyzing the temporal evolution of the strain surrounding a mechanical singularity.
Phase singularities sampled over a short time interval ($ \approx 0.4 \mathrm{s}$) on surface of 3D tissue during dynamics as shown  in Figs.~\ref{fig:spiral3d}-\ref{fig:mechbreakup}:
(a) Single  scroll wave with fibers aligned along $x$-axis.
(b) Single  scroll wave with fibers rotated by $90^{\circ}$
(c) Scroll wave chaos with wave breakup.}
\label{fig:elimination}
\end{figure}

To conclude the results section, we present a method to distinguish between paired electromechanical singularities and unpaired mechanical singularities, which we base on our observations of the behavior of the complex amplitude maps surrounding the mechanical singularities, see Fig.~\ref{fig:2danisospiral}(d) and Fig.~\ref{fig:ionicspiralfiberx}(e).
We exploit that the amplitude of deformation waves is significantly smaller in the vicinity of unpaired mechanical singularities compared to paired electromechanical phase singularities. 
To do so in a consistent way, we first normalize the amplitude map between $0$ and $1$ at each time step.
Then $\oint A(\bm{r})dl$ is calculated around a mechanical phase singular point, where $A$ is the amplitude and $l$ is a closed path around that point. This closed path is one grid square, \ie
\begin{align}
\oint A_{i,j}(\bm{r})dl = A_{i,j} + A_{i+1,j} + A_{i+1,j+1} + A_{i,j+1}
\end{align} 
A singular point is then selected as being paired to an electrical singularity or unpaired depending on whether 
the amplitude is larger or smaller than some threshold (\eg $0.15$).
Fig.~\ref{fig:elimination} shows how paired and unpaired mechanical phase singularities can be discriminated using this filtering method. 
The figure shows the spatiotemporal organization of electrical and mechanical phase singularities on the top surface of the bulk tissue. 
Panel (a) depicts the case of a single spiral and linear transverse fiber alignment or no fiber rotation, c.f. Fig.~\ref{fig:spiral3d}. Panel (b) presents the same case, but with fiber rotation, c.f. Fig.~\ref{fig:mechbreakup}(a).
Lastly, panel (c) depicts the case of multiple scroll waves due to electrical wave break in the tissue with fiber rotation, c.f. Fig.~\ref{fig:mechbreakup}(b).
In this figure, the electrical and mechanical phase singularities emerge on the surface of the bulk tissue (sampled over a short time interval $ \approx 0.4 \mathrm{s}$), and are shown as black circles and red crosses, respectively, and the filtered mechanical phase singularities are shown as blue circles.
One can see that in cases (a) and (b), where there is a single scroll wave and no electrical wave break, the filtering method accurately selects the paired mechanical phase singularities.
In the wave turbulence regime, the method is still reasonably effective at eliminating the unpaired mechanical phase singularities and the reduced accuracy can be attributed to the aforementioned weaker co-localization of paired electrical and mechanical singularities in this regime.
In summary, it is possible to automatically distinguish paired from unpaired mechanical singularities using only information about the temporal evolution of the strain in proximity to a singularity.

\section{Discussion}\label{sec:discussion}

In this study, we showed that electromechanical phase singularities are an integral and universal phenomenon in elastic excitable media and result from the dynamics and coupling between cardiac excitation waves, anisotropic contraction, and tissue strain. 
Electromechanical phase singularities are composed of electrical and mechanical phase singularities, which emerge in the tissue's electrophysiology and elasticity, respectively. 
They describe similar features of the respective dynamics and have a distinct relationship to each other while exhibiting separate characteristics pertaining to the particular physics of excitation waves and tissue strain, respectively.
While we found that electrical and mechanical phase singularities can emerge as co-localized pairs, in which case they equally describe the core of an electrical spiral or scroll wave, we also found that mechanical phase singularities can furthermore form independently from electrical ones.
Our study provides for the first time a systematic and detailed description of the behavior of electromechanical phase singularities in a range of situations, in 2D and 3D media with and without muscle fiber anisotropy and different boundary conditions, with simple and more complex dynamics, such as focal waves, spiral waves and scroll wave chaos, and with fully relaxed analytically derived strain fields versus strain fields obtained in reaction-diffusion-mechanics simulations.
We were particularly interested in the properties of mechanical phase singularities and in the questions of whether they (i) follow a similar organization as electrical ones and (ii) how reliably they can be used to locate the core region of the electrical spiral and scroll waves that underlie cardiac fibrillation \cite{Gray1998,Witkowski1998,Christoph2018}.

Our study demonstrates that mechanical phase singularities are topological defects, which emerge in the strain dynamics of contracting cardiac tissue due to both rotational electrical excitation waves and muscle fiber anisotropy.
Because active contraction is exerted along muscle fibers, the deformation resulting from a particular excitation wave pattern is highly anisotropic, which manifests as a polarized strain field when measuring tissue strain.
The anisotropy and polarization lead to mechanical wavebreaks, which are characterized by mechanical singularities.
Importantly, mechanical singularities can also emerge in electrical regimes, which do not exhibit electrical phase singularities: focal excitation waves can produce mechanical phase singularities in the presence of anisotropy. 
\AM{We should emphasize that in this article we focused on elucidating the mechanism for the creation of mechanical singularities using minimal representations of the electrophysiology and mechanical response. This allowed us to rigorously dissect the problem.
There is no question that our approach neglects the complex heterogeneous structure and geometry of the anatomical heart. However, we believe that our extensive set of simulations clearly shows that mechanical wavebreaks are the origins of these singularities.
}
For instance, in the linearly transverse anisotropic case, an elliptical ring-shaped excitation wave produces two mechanical waves propagating in opposite directions along the fiber direction away from the focus, while the strain vanishes in the perpendicular direction to the fiber direction, see Figs.~\ref{fig:2dpace}(b) and~\ref{fig:planewave}(d,f). 
The elliptical excitation wave and the two resulting mechanical waves are characterized by four mechanical phase singularities, organized in two pairs surrounding the focus, where each pair describes one of the mechanical waves, and the phase singularities indicate the mechanical wavebreak locations, respectively.

In the generic situations that we studied in Figs.~\ref{fig:2disospiral}-\ref{fig:2dpace}, electromechanical phase singularities can be divided into paired or unpaired electrical and mechanical singularities. However, in more complex situations, their relationship can be more complicated.
Paired electrical and mechanical singularities are co-localized, only exist in the presence of spiral and scroll waves, and emerge equally with and without anisotropy, c.f. Figs.~\ref{fig:2disospiral} and~\ref{fig:2danisospiral}.
Paired mechanical singularities originate from the rotation of a strain field that is enslaved to the rotation of an electrical excitation wave, where the pairing occurs close to the tip or core region of the electrical spiral or scroll wave, which is typically characterized by an electrical singularity.
Unpaired mechanical and electrical singularities occur in the presence of anisotropy with both focal and spiral or scroll waves because, in anisotropic tissues, focal waves inherently produce mechanical singularities, and spiral or scroll waves produce additional unpaired mechanical singularities outside of their core regions, as shown in Figs.~\ref{fig:2danisospiral}(b,c),~\ref{fig:spiral3d_small} and~\ref{fig:spiral3d}.
However, during three-dimensional chaotic scroll wave dynamics, we observed neither purely paired nor unpaired, but instead partially paired electromechanical filaments close to the scroll wave's core. The degree of co-alignment of the partially paired filaments varies in space and over time, respectively; see Fig.~\ref{fig:mechbreakup}(c,d). At this point, we can only speculate about the origins of this phenomenon, it could be that scroll wave drift or the complex fiber rotation cause the partial dissociation, or it could be that the local strain field is distorted by neighboring waves and their contractions, and the phenomenon needs further investigation.

From an electrophysiological perspective, mechanical phase singularities may carelessly be disregarded as an ambiguous, even unreliable concept to describe electrical rotors because unpaired mechanical singularities do not co-localize with electrical singularities and could accordingly be interpreted as false positives. Further, paired mechanical singularities are merely proxies for the positions of electrical singularities. 
However, while we demonstrated that it is possible to automatically distinguish paired from unpaired mechanical singularities, see Fig.~\ref{fig:elimination}, therefore demonstrating that the localization of electrical singularities via paired mechanical singularities would be practically feasible,
our research also shows that mechanical singularities are much more than a mere description of electrical phenomena. 
They constitute a more generalized way to characterize electromechanical tissue dynamics because they simultaneously reflect the topology of excitation wave patterns, strain dynamics, and interactions of the excitation with the underlying mechanical substrate.
The spatiotemporal organization of tissue deformation is far more complex than the spatiotemporal organization of electrical waves, and mechanical phase singularities reflect this more complex organization. This is exemplified by spiral and focal excitation waves, each producing a very characteristic strain and mechanical singularity pattern. Mechanical singularities encode both the principal morphology of the excitation wave pattern as well as the underlying organization of muscle fibers and the deformation state of the muscle.
For instance, single high-frequency focal sources produce a complex pattern of unpaired mechanical filaments, see Fig.~\ref{fig:mechbreakup}(e--f), and the filaments are a signature of this particular activation pattern together with a specific underlying muscle fiber organization, and not just spurious topological defects.
To further avoid the impression that mechanical singularities are poor proxies for 
electrical singularities, we would like to emphasize that  
unpaired mechanical singularities are only numerous in a regime exemplified in Figs.~\ref{fig:spiral3d} and~\ref{fig:ionicspiralfiberx} where the tissue size far exceeds the spiral wavelength, which is uncommon during ventricular fibrillation. 
Figs.~\ref{fig:2danisospiral} and~\ref{fig:spiral3d_small} are more representative of an episode of polymorphic ventricular tachycardia or fibrillation with fewer unpaired mechanical singularities when the wavelength of reentrant waves is comparable to the tissue size. 
Therefore, we expect the number of unpaired singularities to be typically comparable to the number of paired singularities during high-frequency cardiac arrhythmias. Unpaired singularities may even be completely absent for slower forms of anatomical tachycardias with large wavelengths.

One limitation of the present study is that it rests on the assumption that the calcium transient, and hence mechanical contraction, is in phase with electrical excitation. This assumption remains valid as long as calcium release from intracellular calcium stores is triggered by calcium entry into the cell via L-type calcium channels following membrane depolarization. The fact that spiral waves can be imaged in heart tissue from an optical mapping of the calcium signal \cite{cherry2008visualization} provides direct evidence that this calcium-induced-calcium-release (CICR) mechanism, which underlies normal physiological function \cite{Bers2002}, can remain operative during reentrant cardiac arrhythmias. However, in other settings, the calcium transient may be suppressed if the activation interval is too short for intracellular stores to refill with calcium during the action potential, or may no longer be in phase with the voltage signal, e.g. if the calcium transient occurs by spontaneous release from intracellular stores instead of CICR following membrane depolarization. In those settings, even paired mechanical singularities may cease to exist. 
\Rev{Another limitation of the present study is that it neglects mechanoelectrical feedback \cite{kohl2003cardiac}. 
This feedback, which is complex and still not completely understood, originates from the fact that stretch can modify passive constitutive properties such as membrane capacitance and electrical coupling and activate mechano-sensitive ion channels. This feedback has been shown to influence properties of excitation waves such as action potential duration restitution, conduction velocity, and to even induce spiral drift \cite{Weise:2013aa}. However, even in the presence of mechanoelectrical feedback, the wave pattern of active force still tracks closely the one of electrical excitation as long as CICR remains operative, e.g. a drifting spiral wave of excitation will induce a drifting pattern of active force. The resulting field of tissue deformation, however, will still exhibit a different spatiotemporal organization including wavebreaks due to the non-locality of elastic interactions. Therefore, we do not expect that   mechanoelectrical feedback would qualitatively change our main finding of the existence of both paired and unpaired mechanical singularities.}

In the future, our findings could play an important role in the interpretation of ultrasound imaging data of the fibrillating heart. 
Mechanical phase singularities could be used to characterize the three-dimensional tissue dynamics during ventricular or atrial fibrillation and could provide estimates of the anchoring sites of electrical spiral or scroll waves deep within the tissue.
To further assess the feasibility of a mechanics-based arrhythmia imaging technique, we aim to study similar electromechanical phenomena beyond the generic setting in physiologically detailed models, explore potential limitations and determine the effect of heterogeneity and tissue anatomy, as mechanics will be substantially altered in fibrotic or scar tissue.
Lastly, we want to better understand the dissociated electromechanical filament dynamics encountered during scroll wave chaos, which is likely caused by the long-range character of elastic forces. This requires addressing the fundamental question of whether the inverse mapping between the observed deformation wave pattern and the excitation-contraction wave pattern causing the deformation, is unique.
In recent work it was shown {\it in silico} that it is possible to compute even complicated electrical excitation wave patterns, such as chaotic scroll waves, from tissue deformation \cite{Lebert2019,Christoph2020}, suggesting that it could also be possible to find a unique mapping between electrical and mechanical filaments.

\acknowledgements{
The authors acknowledge support from the Center for Interdisciplinary Research on Complex Systems at Northeastern University. J.C. and S.L. acknowledge support from the German Center for Cardiovascular Research, partnersite G\"ottingen. J.C. acknowledges support from the Cardiovascular Research Institute at the University of California, San Francisco. S.L. acknowledges support from the Max Planck Society. The simulations benefited from computing time allocation on Northeastern University Discovery Cluster at the Massachusetts Green High Performance Computing Center (MGHPCC).}

\begin{appendices}

\section{Analytical expressions for spiral and target waves}\label{app:spiral}

The active contraction field, $T_a$ for an $n$-turn Archimedean spiral, rotating at angular frequency $\omega$ for isotropic conduction is given by
\begin{align}{}
	& T_a(x,y,t) = \sum_{i=0}^{n}\exp\left\{-\frac{1}{\sigma_s^{2}}\right.\nonumber\\
	&\times \left[\left(x-\frac{(\theta(x,y,t)+2i\pi) \lambda_s \cos(\theta(x,y,t)+\omega t)}{2\pi}\right)^2 \right. \nonumber\\
	& +\left. \left. \left(y-\frac{(\theta(x,y,t)+2i\pi) \lambda_s \sin(\theta(x,y,t)+\omega t)}{2\pi}\right)^2\right]\right\}
	\label{eq:spiralequation}
\end{align}
where
\begin{equation}
\theta(x,y,t)=\arctan\left(\frac{-x\sin\omega t+y\cos\omega t}{x\cos\omega t+y\sin\omega t}\right).
\end{equation}
$\lambda_s=(2\pi c)/\omega$ is the spiral wavelength and $c$ is the conduction velocity. 
$\sigma_s$ represents the thickness of the spiral arms.

For the pacing case, the active contraction field is shaped as a target wave pattern generated by a sequence of $n$ stimuli delivered at $(x=0,y=0)$ with period $T$.
The contraction field of the isotropic pacing case is:

\begin{align}{}
	T_a(x,y,t) = \sum_{i=0}^{{n-1}}\exp\left\{-\left(\frac{\sqrt{x^2+y^2}-r_i(t)}{\sigma_s}\right)^2\right\}
	\label{eq:elipsequation}
\end{align}

where $\sigma_s$ is the wave thickness as in the spiral wave case,
$r_i(t)=c(t-iT)\Theta(t-iT)$ is the radius of the $i^{th}$ concentric wave, $c$ is the conduction velocity, and $\Theta(x)$ is the Heaviside step function defined by $\Theta(x)=1$ for $x\ge 0$ and $\Theta(x)=0$ for $x<0$.

The anisotropic case where fibers are aligned along the  $x-$axis can simply be achieved by rescaling the $x$ coordinate ($x\rightarrow x/\sqrt{D_{\parallel}/D_{\perp}}$) in Eqs.~\eqref{eq:spiralequation} and \eqref{eq:elipsequation}. 
in this rescaling $\sqrt{D_{\parallel}/D_{\perp}}$ is the ratio of the diffusivity parallel and perpendicular to the fiber direction, chosen equal to $\sqrt{5}$ in our simulations.

\section{Reaction-Diffusion Model}\label{app:RDmodel}
The main reaction-diffusion equations to model the electric field has the form:
\begin{align}
	&\partial_{t}U = \nabla\cdot(\mathbf{D}\nabla U)+\tau^{-1}_{U}f(U,v)\tag{\eqref{eq:uequation} revisited}\\
	&\partial_{t}v = \tau^{-1}_{v}g(U,v)\tag{\eqref{eq:vequation} revisited}\\
	&\partial_{t}T_a = \chi(U)(K_tU-T_a)\tag{\eqref{eq:taequation} revisited}
\end{align}

The diffusivity tensor $\mathbf{D}$ is define as 
\begin{align}
	\mathbf{D} = 
	 \begin{bmatrix}
	 D_{11} & D_{12} & 0 \\
	 D_{21} & D_{22} & 0 \\
	 0 & 0 & D_{\perp2} \\
	 \end{bmatrix}
	\label{eq:diffmatrix}
\end{align}
with the matrix elements  
\begin{align}
&D_{11}=D_{\parallel}\cos^2\theta(z)+D_{\perp1}\sin^2\theta(z)\\
&D_{22}=D_{\parallel}\sin^2\theta(z)+D_{\perp1}\cos^2\theta(z)\\
&D_{12}=D_{21}=(D_{\parallel}-D_{\perp1})\cos\theta(z)\sin\theta(z)
\end{align}
where $D_{\parallel}$ is the voltage diffusivity parallel to the fiber axis, $D_{\perp1}$ and $D_{\perp2}$ are the diffusivity perpendicular to this axis in each plane, and $\theta(z)$ is the fiber angle with respect to the $y$-axis in the thickness of the tissue. 
The other functions in Eqs. \eqref{eq:uequation} and \eqref{eq:vequation} are defined as
\begin{align}
	&f(U,v) = -U + \{U^{*}-\mathscr{D}(v)\}h(U)\label{eq:fequation}\\
	&g(U,v) = \mathscr{R}(v)\Theta(U-U_v)-\{1-\Theta(U-U_v)\}v\label{eq:gequation}\\
	&h(U) = \{1-\tanh(U-U_h)\}\frac{U^2}{2}\label{eq:hequation}
\end{align}
 where $\Theta(x)$ is the standard Heaviside step function. 
 Note that the rest state of the membrane corresponds to $U=0$ and $v=0$. The functions $\mathscr{R}(v)$ and $\mathscr{D}(v)$ are the \textit{restitution} an \textit{dispersion} functions respectively and are defined as:

\begin{align}
	&\mathscr{R}(v)=\frac{1-[1-e^{-Re}]v}{1-e^{-Re}}\label{eq:restitution}\\
	&\mathscr{D}(v)=v^M\label{eq:dispersion}
\end{align}
where $Re$ is the parameter that controls the restitution properties. 
Increasing $Re$ makes the slope of the action potential duration restitution curve steeper at the short diastolic interval, thereby promoting spiral/scroll wave breakup. 
The parameter $M$ controls the conduction velocity restitution curve
 and increasing $M$ flattens this curve.
In all of the results that are presented in this article except one case, $Re=0.8$ and $M=10$, which corresponds to a regime where a single spiral or scroll wave propagates and does not break up. 
The only exception is the investigation of a 3D parallelepipedal slab of tissue with rotation anisotropy where $Re=1.0$ produces scroll wave breakups. 
More details about this model and the effect of each parameter can be found in \cite{Karma:1994aa}. 

\begin{figure}[]
\centering
\includegraphics[width=\columnwidth,angle=0,trim={0in 0in 0in 0in}]{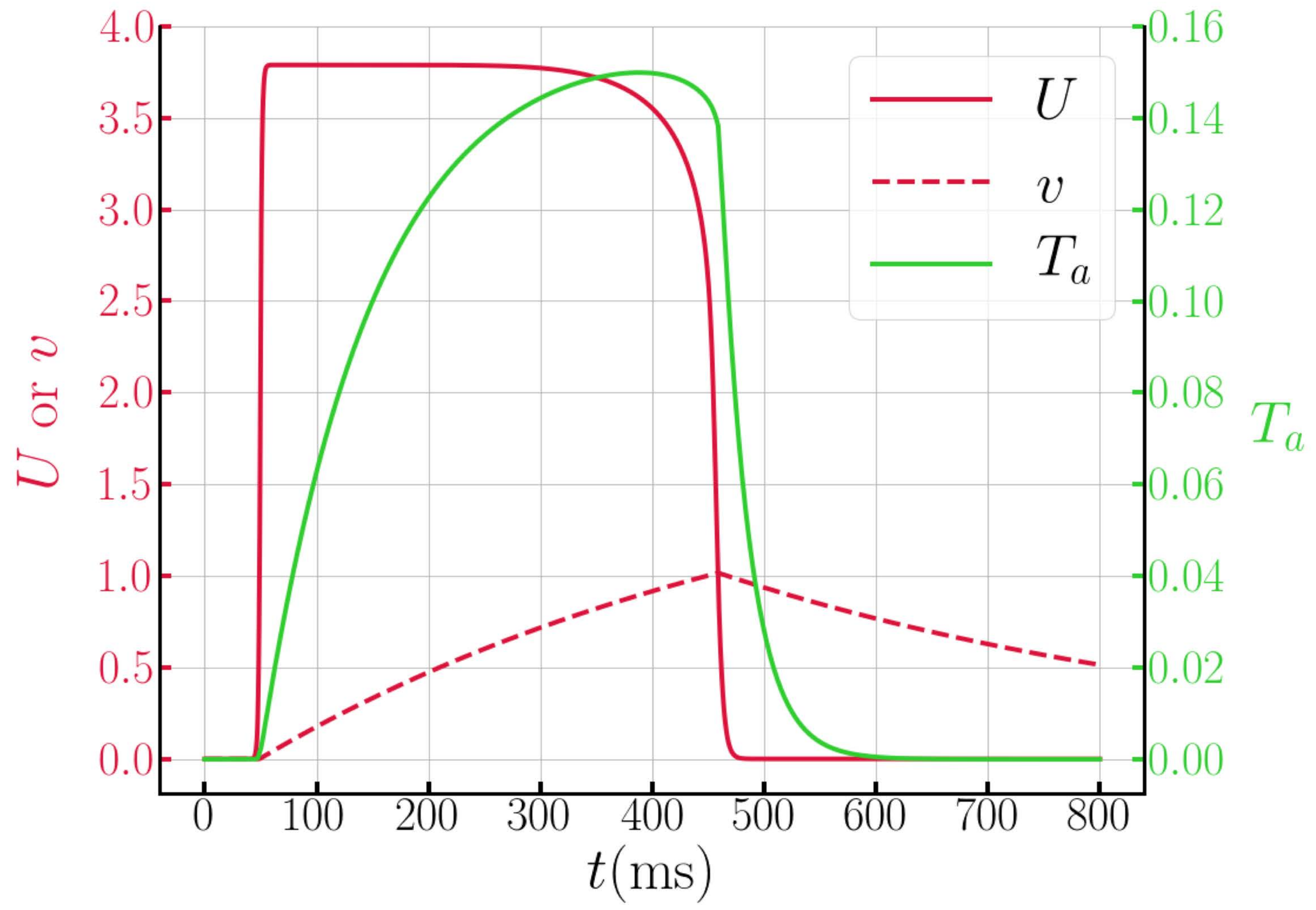}{}
\caption{Characteristic electrical impulse and contraction shapes exhibited by the electrophysiology model described in Sec.~\ref{sec:karma94} and Appendix~\ref{app:RDmodel}. 
In this graph, a 1D representation of the equations is solved using the forward Euler method. 
In this simulation $D=1.1\,\mathrm{cm^2s^{-1}}$, $Re=0.8$, $dx=3\times10^{-2}\,\mathrm{cm}$, and $dt=2.0\times10^{-4}\,\mathrm{s}$. 
Other parameters can be found in Table~\ref{Ta:electroparameters}. 
The left vertical axis corresponds to the variables $U$ or $v$ and the right vertical axis represents the dimensionless magnitude of the active contracting strain $T_a$.
The horizontal axis is time in the units of milliseconds.}
\label{fig:pulse}
\end{figure}

In Eq.~\eqref{eq:taequation}, $\chi(U)$ is a step function that sets the time scale of the contraction period with respect to $U$. It has been set as $\chi(U)=5$ if $U\ge U_h$ otherwise, $\chi(U)=20$. 
The set of parameters that are used in this article are listed in Table~\ref{Ta:electroparameters}. 
A characteristic pulse and contraction structure based on these parameters are shown in Fig.~\ref{fig:pulse}. 
In this figure, a 1D discretization of Eqs.\eqref{eq:uequation}, \eqref{eq:vequation}, and \eqref{eq:taequation} has been solved using the forward Euler method.
One can see that during a single pulse, the voltage surges immediately to its maximum. 
However, the contracting force starts to develop slower than the voltage, and it reaches its maximum when the voltage almost starts to decrease. Even though this model is phenomenological, it reproduces qualitatively the delayed peak contractile force following depolarization. 

\section{Analysis of 1D cable}\label{app:1dcable}

In a human heart with density $\rho\approx 1000\,\mathrm{kgm}^{-3}$ and Young's modulus $E\approx 125\,\mathrm{MPa}$~\cite{Weise:2013aa}, the elastic wave speed can be crudely estimated to be $C\approx 353\,\mathrm{ms^{-1}}$. 
Furthermore, the period of excitation $T$ spans the range of 0.1 to 1.0 seconds with the lower and upper bounds corresponding to an electrical rotor and a normal heartbeat, respectively. 
It follows that the distance $CT$ traveled by elastic waves during a contraction period is much larger than the characteristic size $L$ of the heart of a few centimeters. 
This scale separation ($L\ll CT$) makes it possible to rescale the equations for the computational efficiency of the time integration algorithm.

\subsection{Spatial decay rate of elastic waves} 

To choose our model parameters to be in a regime where $L\ll CT$, we compute the spatial decay rate of elastic waves
starting from the 1D elastodynamics equation with Kelvin-Voigt damping and no external force  
\begin{align}
\rho \partial^{2}_{t} u = E \partial^{2}_{x} u + \eta \partial^{2}_{x} \dot{u},\label{eq:1dcablefull} 
\end{align}
or equivalently  
\begin{align}
\partial^{2}_{t} u = \frac{E}{\rho} \partial^{2}_{x} u + \frac{\eta}{\rho} \partial^{2}_{x} \dot{u}=C^2 \partial^{2}_{x} u + D_e \partial^{2}_{x} \dot{u}\label{eq:1dcable}
\end{align}
Before computing the decay rate, it is useful to first derive expressions for the storage and loss moduli in a setting where a one-dimensional cable is held fixed at one end $(x=0)$ and periodically displaced at the other end $(x=L)$ corresponding to
the boundary condition 
\begin{align}
u(0)=0;~~~~~~u(L)=A\cos(\omega t)
\label{eq:qdcablebc}
\end{align}
where $E$ and $\eta$ are Young's modulus and damping of the system, $C=\sqrt{E/\rho}$ is the characteristic wave speed of the system, $D_e=\eta/\rho$ is the elastic diffusion constant of the system and $L$ is the length of the cable. With $L\ll CT$ ($T=2\pi/\omega$), Eq.~\eqref{eq:1dcablefull} will reduced to 
\begin{align}
E \partial^{2}_{x} u + \eta \partial^{2}_{x} \dot{u}=0
\label{eq:1dcablequasi}
\end{align}
One solution that satisfies this equation has the form of 
\begin{align}
u(x)=\frac{Ax}{L}\cos(\omega t),
\label{eq:1dcabledisp}
\end{align}
which, together with $\rho \partial^{2}_{t} u = \partial_{x}\sigma$, yields
\begin{align}
\sigma=E \partial_{x}u+\eta\partial_{x}\dot{u}=\frac{E A}{L}\{\cos(\omega t)-\tau_R\omega\sin(\omega t)\}
\label{eq:1dcablestress}
\end{align}
\AM{where $\tau_R=\eta/E$ is the retardation time.}
Further using the trigonometry identity $\cos(\omega t +\delta)=\cos(\omega t)-\delta \sin(\omega t )$ in the limit of small $\delta$ we will have 
\begin{align}
\sigma=\frac{E A}{L}\cos(\omega t+ \delta) =  \sigma_0 \cos(\omega t+ \delta)
\label{eq:1dcablestresswithloss}
\end{align}
where $\delta=\eta \omega/E$ and $\sigma_0 = EA/L$. Using Eq.~\eqref{eq:1dcabledisp}, the strain in the cable is
\begin{align}
\epsilon=\partial_{x}u=\frac{A}{L}\cos(\omega t)= \epsilon_0 \cos(\omega t)
\label{eq:1dcablestrain}
\end{align}
where $\epsilon_0 = A/L$. In Eq.~\eqref{eq:1dcablestresswithloss}, $\delta$ is the phase lag between stress and strain. Using Eqs.~\eqref{eq:1dcablestresswithloss} and \eqref{eq:1dcablestrain} the storage and loss moduli are  
\begin{align}
&E_{s} = \frac{\sigma_0}{\epsilon_0} \cos(\delta) \label{eq:storage}\\
&E_{l} = \frac{\sigma_0}{\epsilon_0} \sin(\delta) \label{eq:loss}
\end{align}
Next, to calculate the spatial decay rate of elastic waves,
we substitute a traveling wave solution of the form
\begin{align}
u\approx e^{ikx+i\omega t}
\label{eq:udecay}
\end{align}
into the full elastodynamics equation including inertia (Eq. \eqref{eq:1dcablefull}), which yields
\begin{align}
-\rho \omega^2=-E k^2-i\omega \eta k^2,
\label{eq:1dcabledecay}
\end{align}
or 
\begin{align}
k^2=\frac{\rho \omega^2}{E + i\omega \eta}\rightarrow k=\frac{\omega}{C}\left(1-\frac{1}{2}\frac{i\omega\eta}{E}+\cdots\right)=k_R+ik_I
\label{eq:1}
\end{align}
where the magnitude of the imaginary part of $k$
\begin{align}
|k_I|=\frac{\omega^2\eta}{2E C}\label{eq:kj}
\end{align}
determines the decay rate
\begin{align}
\lambda_d=\frac{1}{|k_I|}=\frac{2E C}{\omega^2\eta}=\frac{CT}{\pi \delta}.
\end{align}

\subsection{Choice of model parameters} 
In our simulations, we set $C=5\,\mathrm{ms^{-1}}$ and the size of the tissue is $4.5\times4.5\times1.2\,\mathrm{cm}^3$.
Therefore, in the limits of $1\,\mathrm{s}$ and $0.1\,\mathrm{s}$ periods, we have:
\begin{align}
&T=1\,\mathrm{s}\rightarrow L=4.5\,\mathrm{cm} \ll CT = 500\,\frac{\mathrm{cm}}{\mathrm{s}}\times1\text{s}=500\,\mathrm{cm} ~~~  \nonumber\\
&T=0.1\,\mathrm{s}\rightarrow L=4.5\,\mathrm{cm} \ll CT = 500\,\frac{\mathrm{cm}}{\mathrm{s}}\times0.1\text{s}=50\,\mathrm{cm} ~~~  \nonumber
\end{align}
It is obvious that the assumption of $L \ll CT$ is valid for both cases.
With $\rho=1000\,\mathrm{kg m}^{-3}$ the Young's modulus of the tissue is $E=C^2\times\rho=25\times10^3\,\mathrm{Nm}^{-2}$.

Now, we need to confirm that with our assumptions, the size of the tissue is much less than the decay rate of the elastic waves.
In our simulations the diffusivity of the tissue is $D=1.1\times10^{-4}\,\mathrm{m}^2\mathrm{s}^{-1}$.
If we set $D_e$ in \eqref{eq:1dcable} to $1.1\times10^{-3}\,\mathrm{m}^2\mathrm{s}^{-1}$ \ie $D_e/D=10$ and further using the relations $D_e=\eta/\rho$, $\delta=\eta\omega/E$, and $\omega = 2\pi/T$, we obtain
\begin{align}
\delta=\frac{D_e \omega}{C^2}=\frac{2.76\times10^{-4}}{T}. 
\label{eq:elasticdiff}
\end{align}
Note that in Eq.~\eqref{eq:elasticdiff} $2.76\times10^{-4}$ has the unit of seconds.
Substituting the excitation period over the range encompassing a normal heart rhythm and fibrillation, we obtain the estimates
\begin{align}
&T=1\mathrm{s}\rightarrow L=4.5\,\mathrm{cm}   \ll \frac{CT}{\pi \delta} \approx2.89\times10^{6}\,\mathrm{cm}  ~~~  \nonumber\\
&T=0.1\mathrm{s}\rightarrow L=4.5\,\mathrm{cm}  \ll \frac{CT}{\pi \delta} \approx2.89\times10^{4}\,\mathrm{cm}  ~~~  \nonumber 
\end{align}
Those estimates are consistent with our assumption that that the distance $CT$ traveled by elastic waves during a contraction period is much larger than the characteristic size $L$ of the heart of a few centimeters. 

\subsection{Adimensionalization} \label{app:nondim}

In this section, we adimensionalize Eq.~\eqref{eq:1dcable} and Eq. \eqref{eq:motion} of the MSM using a characteristic length $L_0=a$ where a is the lattice size, a characteristic time, $\tau_{el}=L_0/C$, and $\rho$. Substitute these characteristic values into Eq.~\eqref{eq:1dcable} and simplifying, one will have:
\begin{align}
\partial^{2}_{\bar{t}} \bar{u} =\partial^{2}_{\bar{x}} \bar{u} + \frac{D_e}{Ca} \partial^{2}_{\bar{x}} \dot{\bar{u}}
\end{align}
In this fashion, the dimensionless characteristic wave speed of the system is unity, and the dimensionless damping of the system is $D_e/(Ca)$.
The corresponding discrete elastodynamics equation for the 1D MSM is
\begin{align}
\mathbf{F}=m\partial_{t}^{2}u+c\partial_{t}u+ku
\label{eq:mass-spring-damper}
\end{align}
where $\mathbf{F}$ is the sum of active and volumetric forces applied to the mass.
Let us adimensionalize Eq.~\eqref{eq:mass-spring-damper} (the same adimensionalization holds in 3D for Eq. \eqref{eq:motion}) by the characteristic length of the lattice size $a$, $E$ the Young's modulus of the tissue, $\tau_{el}$ the characteristic time of the system, and $\rho$ the density of the material, yielding
\begin{align}
\bar{\mathbf{F}}=\frac{\rho a^2}{E \tau_{el}^{2}}\bar{m}\partial_{t}^{2}\bar{u}+\frac{\rho a^2}{E \tau_{el}^{2}}\bar{c}\partial_{t}\bar{u}+\frac{\rho a^2}{E \tau_{el}^{2}}\bar{k}\bar{u}
\label{eq:mass-spring-damper_adimentionalize}
\end{align}
Further substituting $\tau_{el}=L_0/C$ into $(\rho a^2)/(E \tau_{el}^{2})$ will reduce Eq.~\eqref{eq:mass-spring-damper_adimentionalize} to:
\begin{align}
\bar{\mathbf{F}}=\bar{m}\partial_{t}^{2}\bar{u}+\bar{c}\partial_{t}\bar{u}+\bar{k}\bar{u}
\label{eq:mass-spring-damper_adimentionalize_with_tau}
\end{align}
For the electrophysiology model, it is more convenient to adimensionalize time in Eq.~\eqref{eq:1dcablefull} with $\tau=a^2/D$ instead of $\tau_{el}$ where $D$ is the diffusion coefficient of membrane voltage in the electrophysiology model. 
In this case, the dimensionless elastic wave speed is $\tau/\tau_{el}$ and dimensionless damping constant is $D_e/D$. The rescaled continuum elastodynamics equation becomes
\begin{align}
\partial^{2}_{\bar{t}} \bar{u} =\left(\frac{\tau}{\tau_{el}}\right)^2 \partial^{2}_{\bar{x}} \bar{u} + \frac{D_e}{D} \partial^{2}_{\bar{x}} \dot{\bar{u}}.
\end{align}
At the discrete level of the MSM, Eq.~\eqref{eq:mass-spring-damper_adimentionalize} becomes
\begin{align}
\left(\frac{\tau}{\tau_{el}}\right)^2\bar{\mathbf{F}}=\bar{m}\partial_{t}^{2}\bar{u}+\bar{c}\partial_{t}\bar{u}+\left(\frac{\tau}{\tau_{el}}\right)^2\bar{k}\bar{u}.
\label{eq:mass-spring-damper_adimentionalize_with_tau2}
\end{align}
This change in normalizing time becomes important in setting the proper time step to carry out the numerical integration. 
In all the 3D simulations in this article we have set $C=5\,\mathrm{ms^{-1}}$, $dx=3\times10^{-4}\,\mathrm{m}$, $D=1.1\times10^{-4}\,\mathrm{m^2s^{-1}}$, and $D_e=11\times10^{-4}\,\mathrm{m^2s^{-1}}$.
Those choices yield the time constants and adimensionalized elastic wave speed $\bar{C}$ 
\begin{align}
&\tau=\frac{dx^2}{D}=\frac{(3\times10^{-4})^2}{1.1\times10^{-4}}=8.18\times10^{-4}\,\mathrm{s}\nonumber\\
&\tau_{el}=\frac{dx^2}{D_e}=\frac{3\times10^{-4}}{5}=6\times10^{-5}\,\mathrm{s}\nonumber\\
&\bar{C}=\frac{\tau}{\tau_{el}}=\frac{8.181819\times10^{-4}}{6\times10^{-5}}=13.6\nonumber
\end{align}

\section{Mechanical Properties of the MSM}\label{app:msm_properties}

\AMT{To characterize the mechanical properties of the mass-spring model, such as stress--strain relation or storage and loss modulus, we conducted a series of numerical studies.
To approximate one-dimensional settings, we model a long bar on a cubic lattice where all the simulation parameters are presented in Table~\ref{Ta:elastoparameters}.
We used $\rho$, $E$, and $a$ to adimensionalize the equation of motion utilizing equations presented in part C of Appendix~\ref{app:1dcable}.}

\begin{table}[H]
	\caption{Parameters for MSM model used in Appendix~\ref{app:msm_properties}. n.u. stands for no unit.}
	\label{Ta:elastoparameters}
	\begin{ruledtabular}
		\begin{tabular}{l l l}
             $E$ & $125 \times 10^{6}$ & $\mathrm{Pa}$\\
             $\eta$ & $353553$ & $\mathrm{Ns m^{-2}}$\\
             $\rho$ & $1000$ & $\mathrm{kg m^{-3}}$\\
             $C$ & $353.553$  & $\mathrm{m s^{-1}}$\\
             $a$ & $0.01$  & $\mathrm{m}$\\
             $\tau_{el}$ & $2.828 \times 10^{-5}$  & $\mathrm{s}$\\
			 $k$ & $5 \times 10^{5}$ & $\mathrm{N m^{-1}}$\\
			 $c$ & $1414.212$ & $\mathrm{Ns m^{-1}}$\\
			 $dt$ & $0.01128$  & $\mathrm{\mu s}$\\
			 $nx$ & $60$  & $\mathrm{n.u.}$\\
			 $ny$ & $6$  & $\mathrm{n.u.}$\\
			 $nz$ & $6$  & $\mathrm{n.u.}$\\
		\end{tabular}
	\end{ruledtabular}
\end{table}

\begin{figure}[H]
\centering
\includegraphics[width=\columnwidth,angle=0,trim={0in 0in 0in 0in}]{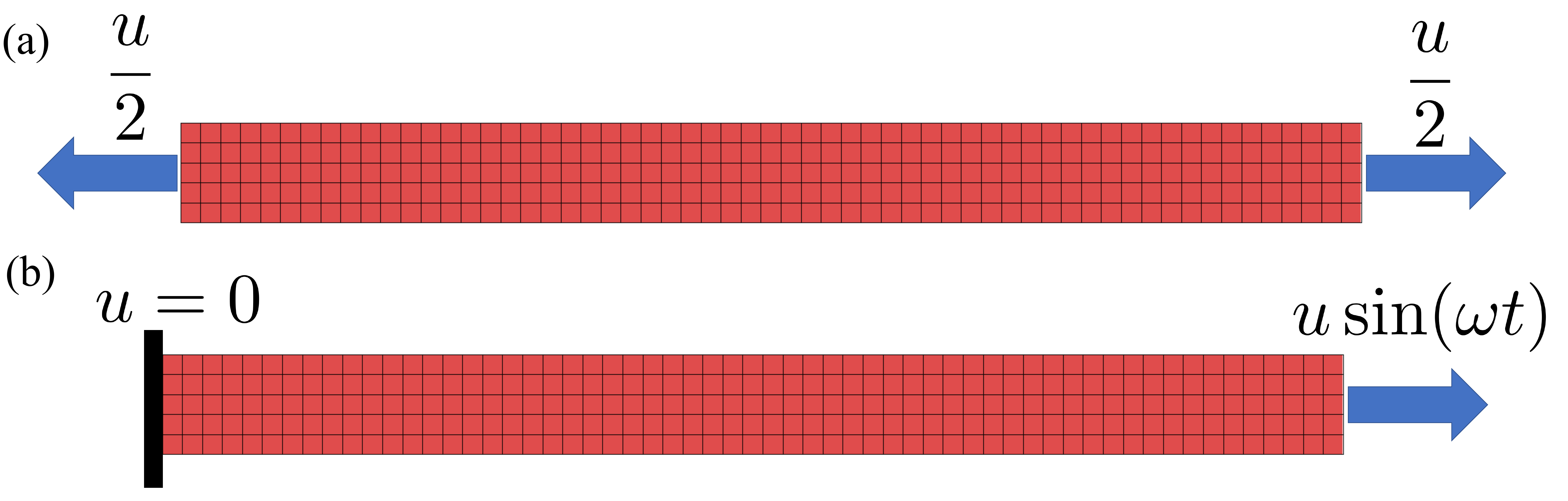}{}
\caption{\AMT{a long bar with aspect ratio $59/5$ has been used to study the mechanical properties of the MSM. (a) Two equal and opposite displacements along the bar's long axis have been applied to the ends of the bar. This loading has been used to find the relation of the Young's Modulus and Poisson ratio with the penalty volumetric pressure and to find the stress--strain relation of the MSM. In this loading case, no boundary condition was introduced because at all times the bar is in force and moment equilibrium. (b) A periodic displacement has been applied to one end of the bar while the other end is kept fixed. This loading has been used to investigate the storage and loss modulus of the MSM.}}
\label{fig:bar}
\end{figure}

\AMT{As shown in Fig.~\ref{fig:bar}(a), a pair of tensile displacements along $x-$axis has been applied on both ends of the bar with traction free boundary conditions on the top and the bottom surfaces.
To calculate Young's modulus and Poisson ratio, for a given imposed strain, we obtain the quasi-static configuration by letting velocities and accelerations of all masses to vanish. 
To speed the simulations, we set the adimensionalized damping constant $D_e/(Ca)=100$ where $D_e=\eta/\rho$ is the elastic diffusivity and $\eta$ is the damping ratio.
Note that the small time step is because of the large damping we used for the simulations.} 

\begin{figure}[H]
\centering
\includegraphics[width=\columnwidth,angle=0,trim={0in 0in 0in 0in}]{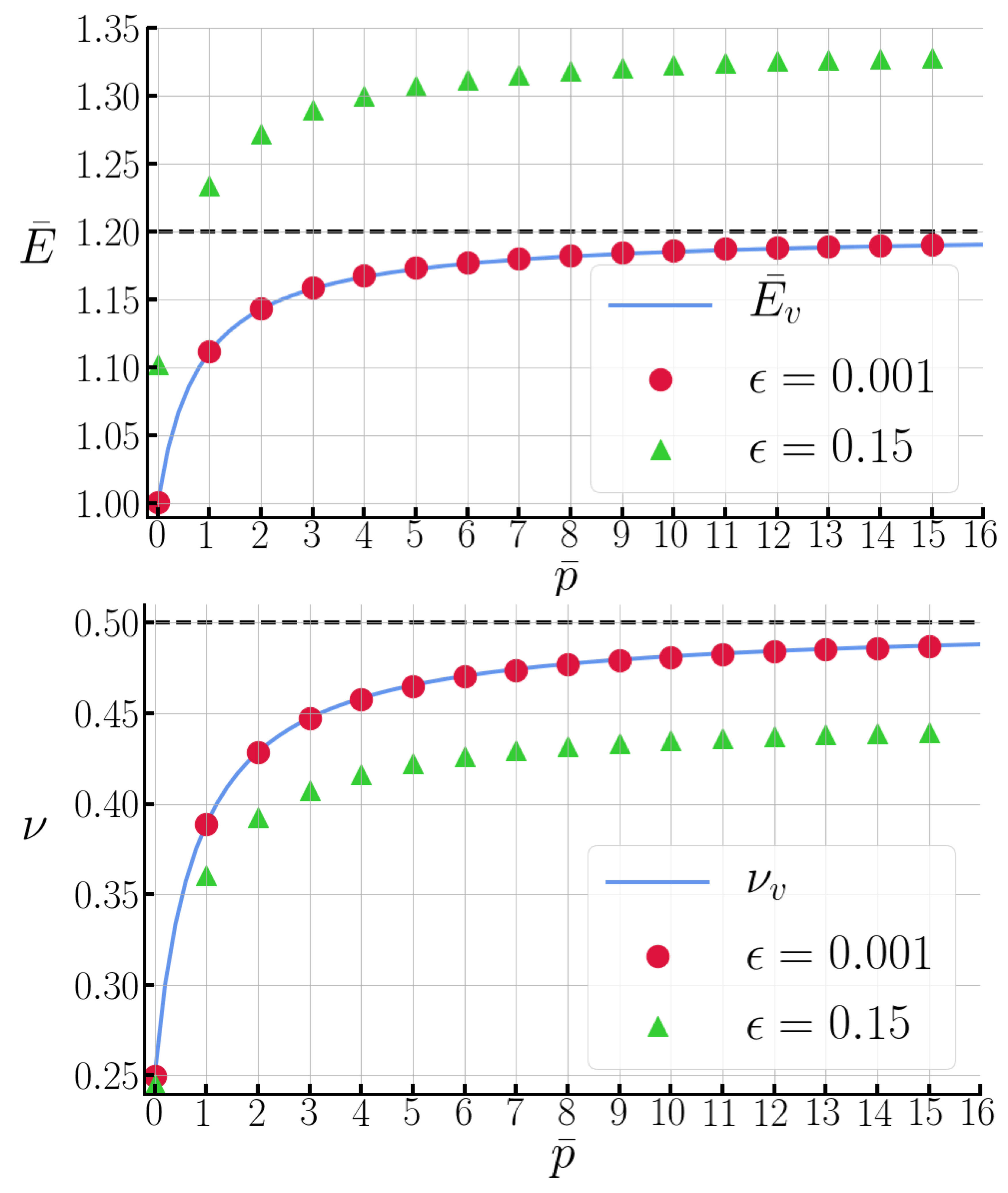}{}
\caption{\AMT{(top) Young's modulus and (bottom) Poisson's ratio of the MSM at small strain $\epsilon=0.1\%$ and large strain $\epsilon=15\%$ as a function of dimensionless volumetric pressure $\bar{p}=p/E$. $\bar{E}_v$ is the dimensionless Young's Modulus which is equal to $\bar{E}_v=E_v/E$.
$E_v$ is from Eq.~\ref{eq:youngvol} and $\nu_v$ is from Eq.~\ref{eq:nuvol}.}}
\label{fig:MSM_e_nu}
\end{figure}

\AMT{First, we investigated the effect of the volume conservation penalty pressure on the elastic properties of the MSM. 
The analytic equations for these relations for infinitesimal strains have been presented in Eqs.~\ref{eq:youngvol} and~\ref{eq:nuvol}.
We studied these properties of the MSM for strains $\epsilon_{xx}=u_x/L=0.001,\,0.15$. 
The former represents the infinitesimal strain elasticity limit and the latter can be considered as the large deformation limit.
The $0.15$ strain limit was chosen because, during a traveling electrical wave, cardiac cells experience about 15\% compression strain along their fibers~\cite{Weise:2013aa}.
In these simulations, stresses are calculated by adding all the reaction forces at one end and dividing it by the engineering cross-section ($5a\times5a$).
Using the calculated stress and applied strain, the Young's Modulus of the MSM is calculated as $E=\sigma_{xx}/\epsilon_{xx}$. 
In addition, we calculated the Poisson's ratio buy finding the ratio between the transverse strain at the middle of the bar and the applied longitudinal strain \ie $\nu=-\epsilon_{yy}/\epsilon_{xx}$. 
Results as a function of the volumetric pressure are presented in Fig.~\ref{fig:MSM_e_nu}.}

\AMT{In the small strain limit $\epsilon=0.1\%$, the simulation results follow the analytical equations of \ref{eq:youngvol} and \ref{eq:nuvol} perfectly.
However, for large strain cases, due to geometric nonlinearities, the MSM exhibits a hyperelastic behavior with higher Young's modulus and lower Poisson's ratio.
In our simulations for the heart tissue, we assume $\bar{p}=p/E=3$ that yields the Poisson's ratio range between $0.41$ and $0.45$ for the strain range between $0.1\%$ and $15\%$.}

\begin{figure}[H]
\centering
\includegraphics[width=\columnwidth,angle=0,trim={0in 0in 0in 0in}]{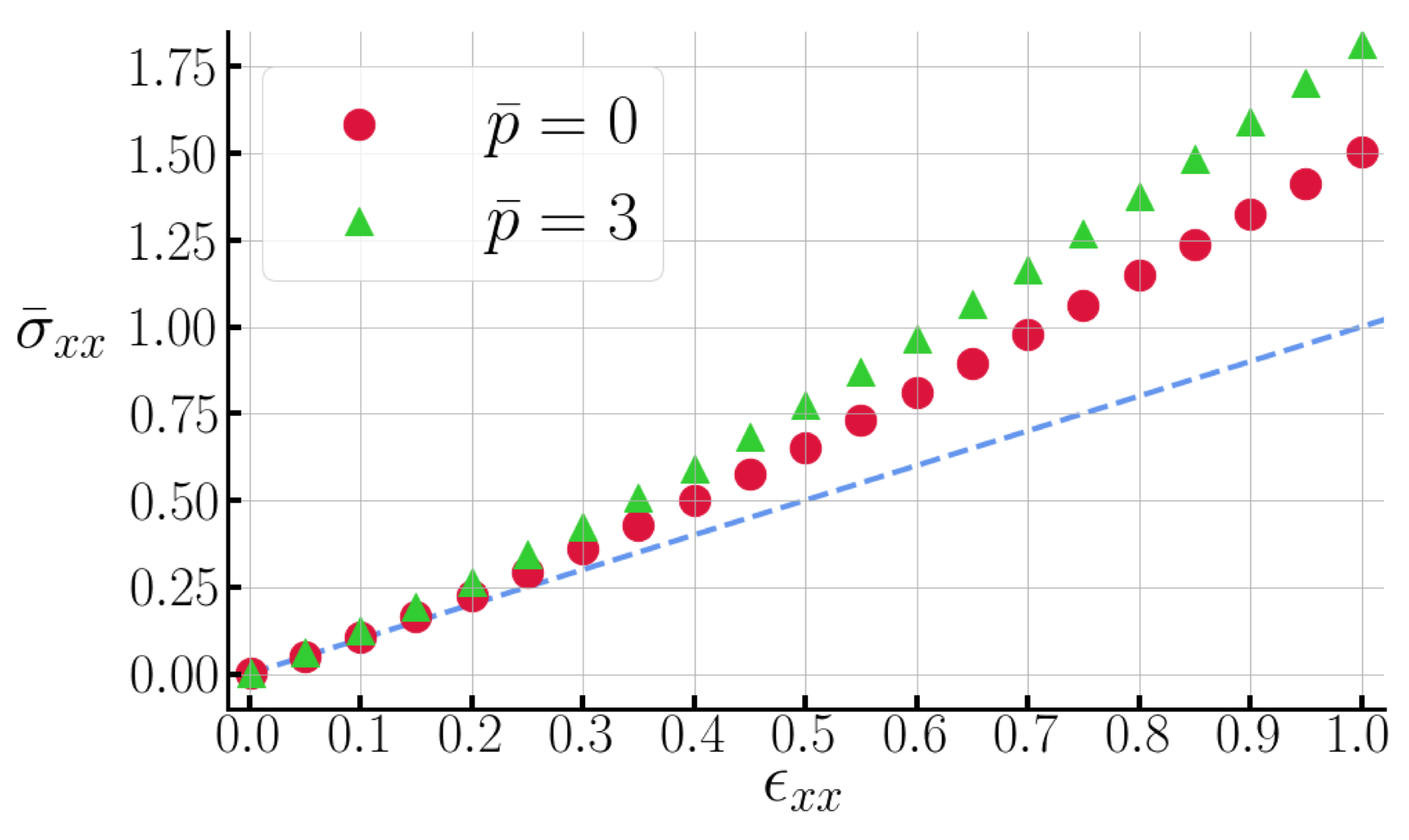}{}
\caption{\AMT{Stress-strain curve of the MSM model. $\bar{p}$ represents the dimensionless volumetric pressure.
Including the volumetric pressure will increase the hyperelasticity.}}
\label{fig:MSMstressstrain}
\end{figure}

\AMT{To study the stress-strain relation of the MSM, 
we incrementally increased the applied displacement (strain) while volumetric penalty pressure was kept constant at a given values $0\leq\bar{p}\leq3$. 
The results, depicted in Fig.~\ref{fig:MSMstressstrain}, clearly show that the presented MSM has a hyperelastic response, where increasing the volumetric penalty pressure will increase the nonlinearity of the model.}

\begin{figure}[H]
\centering
\includegraphics[width=\columnwidth,angle=0,trim={0in 0in 0in 0in}]{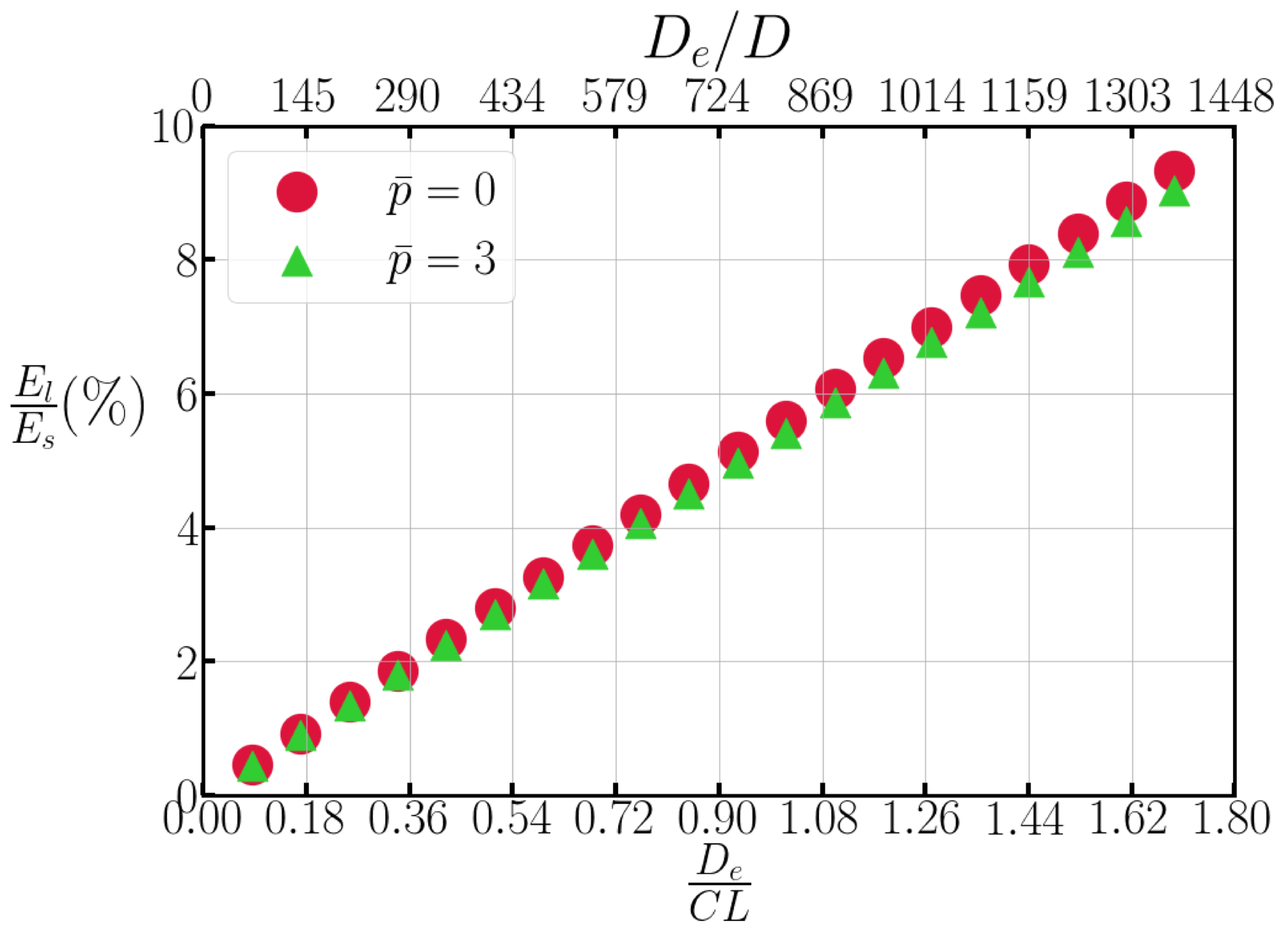}{}
\caption{\AMT{The ratio between loss and storage modulus as a function of dimension less elastic diffusion. $D_e$ is $\eta/\rho$ where $\eta$ is the damping of the system and $\rho$ is the density of the material. $C=\sqrt{E/\rho}$ is the characteristic elastic wave speed and $L$ is the length of the system (not the lattice dimension). The top $x-$axis shows the ratio between $D_e$ and the diffusion coefficient $D=1.1\,\mathrm{cm^2/s}$ in the electrophysiology model. The relation between $E_l/E_s$ and damping of the system is linear in $0.1\%$ strain and at zero damping, loss modulus is zero. The penalty volumetric pressure $p$ slightly decrease the $E_l/E_s$ ratio in high damping constants.}}
\label{fig:MSMloss}
\end{figure}

\AMT{Finally, in the small deformation limit we studied the viscoelastic properties of the MSM. We applied a sinusoidal displacement (strain) with $1\,\mathrm{Hz}$ frequency at $x=L$ while the degrees of freedom at $x=0$ were fixed. We then calculated the stresses from the reaction forces at $x=L$ (see Fig.~\ref{fig:bar}b). 
We define the mechanical transfer function (the complex modulus of the system) as $\tilde{E}=\tilde{\sigma}_{xx}/\tilde{\epsilon}_{xx}$ where $\tilde{\sigma}_{xx}$ and $\tilde{\epsilon}_{xx}$ are Fourier transformed axial stress and strain, respectively. 
We then define the storage modulus as the real part  $E_s=\mathrm{Re}(\tilde{E})$ and the loss modulus as the imaginary part $E_l=\mathrm{Im}(\tilde{E})$ of the complex modulus. 
In a viscoelastic material, the ratio between the loss and storage modulus is a measure of damping in the material. 
This ratio as a function of dimensionless elastic diffusion is shown in Fig~\ref{fig:MSMloss}. 
The elastic diffusion can be defined as the damping coefficient divided by the density of the material (see Appendix \ref{app:1dcable} part C).
As expected, with zero damping there is no loss modulus, and the ratio between loss and storage modulus change linearly with the damping of the system.
The effect of volumetric penalty pressure is not significant.
Details are discussed in Appendix~\ref{app:1dcable}.}

\end{appendices}

\newpage
\section*{References}
%

\end{document}